\documentclass[journal]{IEEEtran}

\usepackage{cite}
\usepackage{amsmath,amssymb,amsfonts}
\usepackage{algorithmic}
\usepackage{graphicx,caption,subcaption}
\usepackage{textcomp}
\usepackage{bbm}
\usepackage{booktabs} 
\usepackage{xcolor,soul,comment} 

\DeclareMathOperator*{\argmax}{\arg\max}

\begin{document}
%
\title{Exposing Fake Images \\ with Forensic Similarity Graphs}

\author{Owen~Mayer,~\IEEEmembership{Member,~IEEE,}
        and~Matthew~C.~Stamm,~\IEEEmembership{Member,~IEEE}
\thanks{O. Mayer and M.C. Stamm are with the Department
of Electrical and Computer Engineering, Drexel Univeristy, Philadelphia,
PA, 19104 USA e-mail: om82@drexel.edu}
\thanks{This material is based upon work supported by the National Science Foundation under Grant No. 1553610. Any opinions, findings, and conclusions or recommendations expressed in this material are those of the authors and do not necessarily reflect the views of the National Science Foundation.}
}


\maketitle

\begin{abstract}
%

We propose new image forgery detection and localization algorithms by recasting these problems as graph-based community detection problems.
To do this, we introduce a novel abstract, graph-based representation of an image, which we call the Forensic Similarity Graph, that captures key forensic relationships among regions in the image. In this representation, small image patches are represented by graph vertices with edges assigned according to the forensic similarity between patches. Localized tampering introduces unique structure into this graph, which aligns with a concept called ``community structure'' in graph-theory literature. 
In the Forensic Similarity Graph, communities correspond to the tampered and unaltered regions in the image. As a result, forgery detection is performed by identifying whether multiple communities exist, and forgery localization is performed by partitioning these communities. 
We present two community detection techniques, adapted from literature, to detect and localize image forgeries. We experimentally show that our proposed community detection methods outperform existing state-of-the-art forgery detection and localization methods, which do not capture such community structure.

\end{abstract}

\begin{IEEEkeywords}
Multimedia Forensics, Forgery Detection, Deep Learning, Community Detection
\end{IEEEkeywords}


\newcommand{\fsim}[2]{S\left({#1},{#2}\right)}

\section{Introduction}

\IEEEPARstart{D}{etermining} the authenticity of digital multimedia is an important social problem. Many important institutions rely upon truthful multimedia content, including news services, law firms, and intelligence agencies. As a result, multimedia forensics researchers have developed a variety of approaches to inspect images for evidence of forgery~\cite{stamm2013information}. Many of these approaches operate by directly identifying traces of manipulation, such as resampling~\cite{popescu2005exposing}, double JPEG compression~\cite{bianchi2012image,farid2009exposing,ye2007detecting}, or contrast enhancement~\cite{stamm2010forensic}. Several approaches work by identifying physical and lighting inconsistencies~\cite{johnson2005exposing, matern2019gradient}. Other approaches have found significant success by identifying inconsistencies in imaging traces such as lens aberrations~\cite{johnson2006exposing,mayer2018lca}, sensor noise~\cite{lukavs2006detecting,chen2007imaging} and general residual-based features~\cite{cozzolino2015splicebuster,cozzolino2016single}. 
More recently, deep learning based approaches have been found to be very effective for detecting and localizing image forgeries~\cite{verdoliva2020media}. Several techniques have been proposed, including methods that target double JPEG compression~\cite{barni2017aligned,amerini2017localization,wang2016double}, tampering boundaries~\cite{salloum2018image,bappy2019hybrid}, in-painting~\cite{li2019localization}, face-warping~\cite{wang2019detecting}, general manipulations~\cite{wu2019mantra,salloum2018image,mayer2019similarity,zhou2018learning}, and source inconsistencies~\cite{bondi2017cvprw,huh2018forensics,cozzolino2018camera,cozzolino2018noiseprint,mayer2018similarity,mayer2019similarity}.
\looseness=-1


A recent deep-learning-based forensics approach called ``Forensic Similarity'' has been shown to be a promising technique for forgery detection and localization~\cite{mayer2018similarity,mayer2019similarity}. Earlier work in~\cite{mayer2018similarity} demonstrated utility in forgery localization, and later work~\cite{mayer2019similarity} proposed a technique using Forensic Similarity that outperformed prior art in forgery detection. Forensic Similarity is a technique that maps two small image patches to a score indicating whether the two patches contain the same or different forensic traces. These forensic traces are related to the source camera model and/or processing history of the image. Importantly, Forensic Similarity is effective even on ``unknown" forensic traces that were not used to train the system. By identifying differences in forensic traces within an image, falsified images are exposed~\cite{mayer2018similarity,mayer2019similarity}.

While Forensic Similarity has shown encouraging promise for forgery analysis, existing methods have several drawbacks. First, previously proposed forgery localization techniques based on Forensic Similarity require the selection of a reference patch in a region known to be unaltered. This type of approach is sensitive to the selection of reference patch, and is an unrealistic scenario for a forensic investigator, who may not know \textit{a priori} which regions are unaltered. Second, the forgery detection approach using Forensic Similarity in~\cite{mayer2019similarity} uses a heuristic detection measure. While the heuristic approach achieved high forgery detection rates, performance can be significantly improved by developing more powerful approaches.\looseness=-1


To address these problems, in this work we build off of prior Forensic Similarity research to more accurately detect image forgeries and localize the falsified image regions. To do this, we propose a new multimedia forensics concept called the Forensic Similarity Graph, which is an abstract graph-based representation of an image that captures important forensics relationships among all image regions. Briefly, to do this we first sample small patches from an image and represent each patch as vertex of a graph. Then, we assign edge weights between vertices according to the forensic similarity value between each corresponding pair of patches.
Localized image tampering (e.g. splicing or localized airbrushing) introduces particular structures, called communities, into this Forensic Graph Representation. We propose two techniques to identify and partition these communities, adapting community detection techniques from literature~\cite{fortunato2010community}. Using our proposed technique we expose localized image tampering with high accuracy, and do so without requiring a reference patches to be selected.\looseness=-1

The remaining sections of this paper are organized as follows. In Sec.~\ref{sec:related}, we discuss related works on forgery detection and localization, as well as related community detection literature. In Sec.~\ref{sec:approach} we first introduce the concept of the Forensic Similarity Graph, and describe how to compute it in an image. In this section we also show how localized tampering introduces identifiable structures, called ``communities," into the Forensic Similarity Graph.
In Sec.~\ref{sec:approach:ssec:detection}, we propose two community detection techniques which are used to identify and partition these community structures. We show how these techniques are used to detect forged images and, in tampered images, localize the tampered regions. Furthermore, we describe how the community partitions can be converted to pixel-level forgery localization decisions. 
%
%
Finally, in Sec.~\ref{sec:detection_experiments} and Sec.~\ref{sec:localization_experiments}, we conduct a series of experiments testing the efficacy of our proposed approach. Experiments show that our proposed approach outperforms naive approaches that do not consider community structure, improving upon heuristics proposed in prior art. We also show that our approach outperforms prior-art localization approaches on publicly available benchmark databases.

\section{Background and Related Work}
\label{sec:related}
\subsection{Image Forgery Detection and Localization}


Researchers have developed a number of heurisitc and hand derived features for image forgery analysis~\cite{popescu2005exposing,bianchi2012image,farid2009exposing,
ye2007detecting,stamm2010forensic,mayer2018lca,lukavs2006detecting,
chen2007imaging,cozzolino2015splicebuster,johnson2006exposing}.
More recently, research has shown that deep learning based approaches are able to achieve improved performance in forgery detection and localization~\cite{verdoliva2020media}. Work by Barni et al.~\cite{barni2017aligned}, Amerini et al.~\cite{amerini2017localization}, and Wang et al.~\cite{wang2016double}, trained deep learning systems to localize image areas with double JPEG compression artifacts indicative of tampering. Research by Salloum et al.~\cite{salloum2018image} developed a multi-task fully convolutional network, which improved forgery localization by training a branch to learn boundaries of spliced regions. Bappy et al.~\cite{bappy2019hybrid} proposed a system to learn splice-pristine region boundary transition using an LSTM architecture. Work by Wu et al.~\cite{wu2019mantra} also proposed a fully convolutional network, but trained on a targeted set of 385 manipulations. 

The above mentioned deep-learning based techniques are trained to identify a targeted set of forgery features. Recently, researchers have discovered that techniques which identify \textit{inconsistencies} in forensic features~\cite{cozzolino2015splicebuster,mayer2019similarity}, and in particular features related to image source~\cite{bondi2017cvprw,huh2018forensics,cozzolino2018noiseprint,mayer2018similarity}, can be even more successful. Since it is not feasible to train a system on all possible and real-world manipulations, these systems instead detect inconsistencies and anomalies in forensic features indicative of splicing.

In early work by Bondi et al. a convolutional neural network (CNN) was used to generate deep-feature representations of an image's source camera-model, and then detected and localized image forgeries via an iterative k-means clustering approach~\cite{bondi2017cvprw}. Subsequently, our work in~\cite{mayer2018similarity} showed that the similarity of camera-model related forensic traces can be directly measured using a CNN-based siamese network, which can be used to localize image forgeries. This idea was further refined by our research in~\cite{mayer2019similarity}, where we proposed a more formal definition of Forensic Similarity, which is the quantifiable measure of similarity of forensic traces related to the source and/or processing between two images patches, and refined the siamese network technique.\looseness=-1

Deep learning forensics research by Cozzolino et al.~\cite{cozzolino2018noiseprint,cozzolino2018camera} proposed a CNN that transforms an image to highlight artifacts associated with camera-specific traces. Inconsistencies in the resulting fingerprint map were then used to identify forged regions. Huh et al. developed a deep-learning technique to create a ``consistency map'' for forged images~\cite{huh2018forensics}. In this consistency map approach, regions of the image are highlighted which contain predictions of EXIF-based metadata that are inconsistent with the majority of the image. Huh et al. showed that taking spatial average of the consistency map and comparing to a threshold can be used to perform forgery detection.\looseness=-1

Multimedia forensics techniques explicitly draw the distinction between forgery detection and forgery localization~\cite{bondi2017cvprw,huh2018forensics}. Forgery \textit{detection} methods are those that determine if an image has been tampered or is alternatively unaltered. Forgery \textit{localization} methods are those that, given a forged image, determine which regions of an image have been tampered. While these are distinctly different problems, the underlying mechanisms for analysis are similar and, as a result, are often studied in conjunction. In this paper, we propose techniques for both forgery detection and forgery localization.

The work in this paper directly builds from our prior research on Forensic Similarity in~\cite{mayer2019similarity}. In~\cite{mayer2019similarity}, a deep-learning system was designed to input two image patches and output a score indicating forensic similarity. The system was composed of a two CNNs in a hard-sharing configuration, with the CNN architecture based upon the CNN architecture designed in~\cite{bayar2018constrained}, which was shown to be generalization to many forensic tasks in~\cite{mayer2018unified}. These CNNs act as feature extractors, which feed a pair of features, one for each image patch, into a shallow three-layer neural ``similarity network" which has a final layer composed of a single neuron, which outputs a score indicating similarity. This system is then trained end-to-end with pairs of image patches and labels associated with whether they contain the same or different forensic traces.

In~\cite{mayer2019similarity}, we showed that tampered regions of an image can be highlighted by selecting reference patch, and highlighting the patches in the image that contain different forensic traces. Provided a patch in the unaltered region of the image was selected, the tampered regions are highlighted. Furthermore, we showed state-of-the-art forgery detection accuracy by computing the forensic similarity values between all patches, computing the mean of these values, and comparing that mean to a threshold. In unaltered images, this value would be high indicating all patches were forensically similar. Alternatively in tampered images, this value would be low indicating some patches were forensically dissimilar.

There are, however, drawbacks to existing forgery detection and localization approaches using Forensic Similarity. First, the forgery localization approach require a selection of a reference patch in the image. This is problematic since the investigator may not have knowledge of which regions of the image are unaltered, and the poor selection of a patch may lead to erroneous results. Second, the forgery detection approach, which utilizes the mean forensic similarity value of the image, does not adequately consider the complex forensic relationships that occur in a tampered image. For example, a large tampered region will be self-similar within the region, but dissimilar to the unaltered part of the image. This type of relationship is lost by averaging all similarity values. Furthermore, by utilizing the mean similarity value, the detection statistic becomes inherently tied to the size of the forged region. However, it is important to detect small forgeries with high accuracy. This problem also occurs in the forgery detection approach by Huh et al.~\cite{huh2018forensics}, which utilizes the spatial average of their EXIF-based consistency map to detect forgeries. 
In this paper, we propose techniques that address these drawbacks.


\subsection{Community Detection in Graphs}
\label{sec:background:ssec:community}
In this paper, we propose a graph-based representation of the image which is then analyzed for forgery. 
A graph $G=(V,E)$ is defined by a set of elements called ``vertices'' $V$, with ``edges'' $E \subseteq V^2$ that connect pairs of vertices\cite{fortunato2010community,diestel2005graph}. In this work, we propose a graph-based representation of images, with small image regions represented by vertices and edges assigned according to the similarity of their forensic traces. \textit{Communities} are subsets of vertices with dense edges within their own community~\cite{fortunato2010community}. In Sec.~\ref{sec:approach}, we show that localized tampering introduces communities into this graph representation. As a result, localized tampering is exposed via detection of multiple communities in the graph, and the tampered regions are localized by partitioning the vertices according to their respective communities. 

Researchers have developed a variety of techniques to detect and partition communities in general graphs. These techniques are referred to as ``community detection'' algorithms~\cite{fortunato2010community}. Early techniques included divisive techniques, which iteratively remove edges from the graph until communities form as disconnected subgraphs~\cite{girvan2002community}. Other techniques optimize community partitions based on community quality measures such as modularity~\cite{blondel2008fast,newman2004fast,clauset2004finding}, label propagation~\cite{raghavan2007near}, the minimum energy state of the spin-glass model~\cite{reichardt2006statistical}, and probability derived from random walks~\cite{pons2005computing}. Some clustering algorithms can also be framed as community detection techniques, including spectral clustering~\cite{von2007tutorial} and extensions of k-means clustering~\cite{rattigan2007graph}. In the multimedia forensics field, spectral clustering has been applied to the image phylogeny problem~\cite{oikawa2015manifold}.

While there are many approaches to community detection in graphs, we focus on just two for this paper. However, we take care to provide a framework that is readily extended to any community detection technique, allowing for future research to take advantage of improvements in community detection algorithms. Namely the two techniques we focus on are: 1) Modularity Optimization~\cite{newman2004fast} and 2) Spectral Clustering~\cite{von2007tutorial}. These techniques are chosen since they have been shown to be powerful and popular techniques in literature~\cite{fortunato2010community,von2007tutorial}.

\subsubsection{Spectral Clustering}
Spectral clustering is a technique for partitioning graphs into clusters~\cite{fortunato2010community,von2007tutorial,chung1996spectral}, which leverages properties of the graph Laplacian matrix to determine community structures.

The graph Laplacian matrix, $L$, is defined as
\begin{equation}
L = D-W,
\label{eq:graph_laplacian}
\end{equation}
where $W$ is the edge weight matrix with elements $W_{i,j}$ assigned according to the similarity of vertices $i$ and $j$, and $D$ is the degree matrix with values
\begin{equation}
D_{ii} = \sum_j W_{ij}
\end{equation} 
on the diagonal, and zeroes off-diagonal. 
Additionally, we define the \textit{normalized} graph Laplacian 
\begin{equation}
L_{norm} = D^{-\frac{1}{2}}LD^{-\frac{1}{2}},
\end{equation}
which regularizes the matrix~\cite{chung1996spectral}.

The graph Laplacian has a number of special properties~\cite{chung1996spectral}. In particular, it has spectrum of non-negative real valued eigenvalues $0= \lambda_1 \leq \lambda_2 \leq \ldots \leq \lambda_n$. These eigenvalues a have property such that the multiplicity of $\lambda=0$ is equal to the number of disconnected communities in the graph. Furthermore, the eigenspace of $0$ is spanned by the indicator vectors $\mathbbm{1}_{A_1},...,\mathbbm{1}_{A_k}$ of those k communities, where $A_i$ is the vertex membership of the $i^{th}$ community. For the normalized graph Laplacian, it this eigenspace spanned by $D^{\frac{1}{2}}\mathbbm{1}_{A_1}$~\cite{von2007tutorial,chung1996spectral}. We leverage these properties for detecting forgeries and localizing the forged regions, described in Sec.~\ref{sec:approach:ssec:detection}.
\subsubsection{Modularity Optimization}
Modularity, $Q$, i an index for the quality of how well a graph has been partitioned into communities~\cite{fortunato2010community}, defined as:
\begin{equation}
Q = \frac{1}{4m}\sum_{i,j}\left(W_{ij}-\frac{d_id_j}{2m}\right)\mathbbm{1}(c_i = c_j)
\end{equation}
where $m$ is the sum of edges in the graph, $W$ is the edge weight matrix, $d_i$ is the weighted degree of vertex $i$, and the indicator function $\mathbbm{1}(c_i = c_j)$ is 1 when both vertex $i$ and $j$ are assigned to be members of the the same community. 
The term $\frac{d_id_j}{2m}$ is related to the expected weight of a randomly occurring connection between vertices $i$ and $j$. Intuitively, if two vertices in the same community have a low connection expectation, but have a high weight between them, then modularity is increased.

Higher values of modularity indicate that community structure exists, and that a good partitioning has been found~\cite{fortunato2010community,newman2004finding,girvan2002community}. Modularity in the case of a single community is zero.

Modularity Optimization is a family of techniques used to determine the underlying community memberships in a graph. These techniques operate by assigning vertices to communities such that modularity is maximized, 
\begin{equation}
Q_{opt} = \max_{\{c_1, \ldots, c_n \}} \frac{1}{4m}\sum_{i,j}\left(W_{ij}-\frac{d_id_j}{2m}\right)\mathbbm{1}(c_i = c_j),
\label{eq:mod_optimization}
\end{equation}
where $Q_{opt}$ is the maximum found modularity value in the graph. There have been several proposed methods for optimizing for modularity, including divisive and agglomerative algorithms. Pos and Latapy proposed an agglomerative approach based on random walks~\cite{pons2005computing}. Blondel et al. proposed a optimization approach that iteratively applies local changes to community memberships until a maxima is found~\cite{blondel2008fast}. Reichart and Bornholdt posed modularity optmization as a simulated annealing problem, with good results~\cite{reichardt2006statistical}. In this paper we utilize the "fast greedy" method proposed by Clauset et. al~\cite{clauset2004finding}, which uses a hierarchical agglomeration approach, and achieves similar results of other methods on a reasonable time scale. 


\newlength{\imht}
\setlength{\imht}{1.25in}

\newlength{\subht}
\setlength{\subht}{1.45in}


\begin{figure}
\null\hfill
\begin{subfigure}[t]{0.45\linewidth}
\centering
\includegraphics[width=\linewidth]{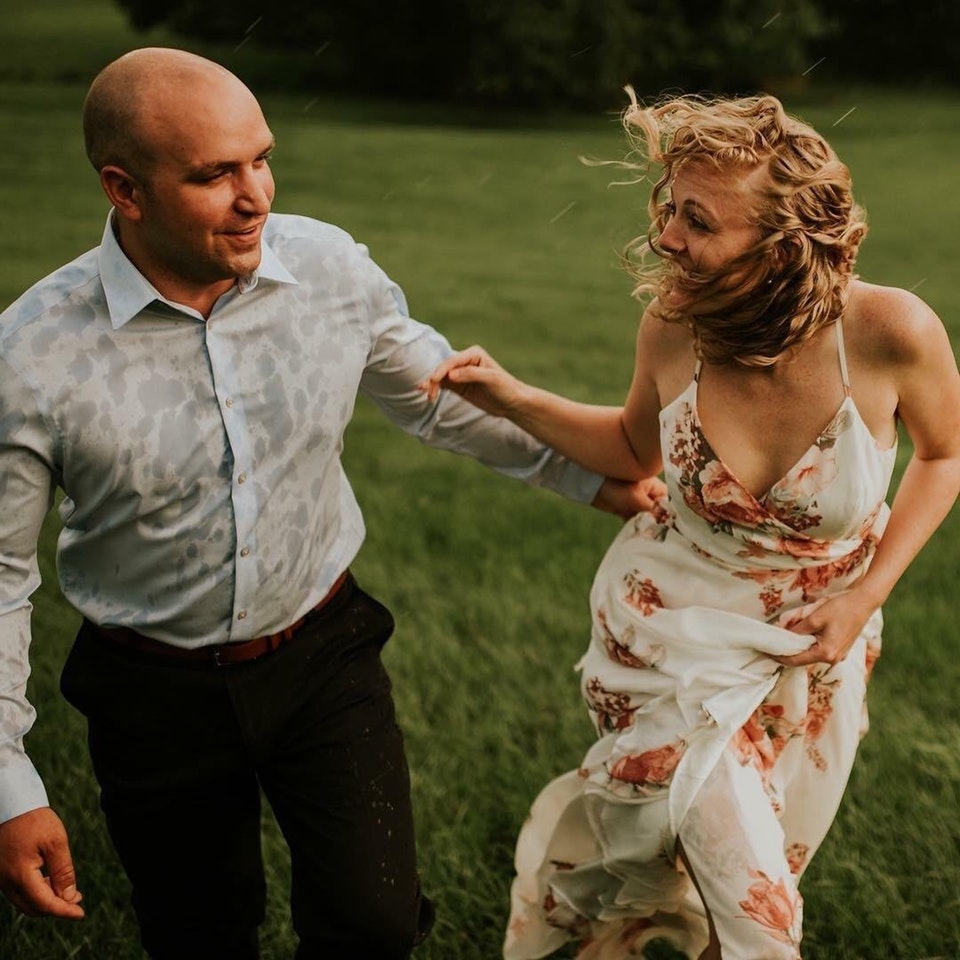}
\caption{Original Image}
\end{subfigure}
\hfill
\begin{subfigure}[t]{0.46\linewidth}
\centering
\includegraphics[width=\linewidth]{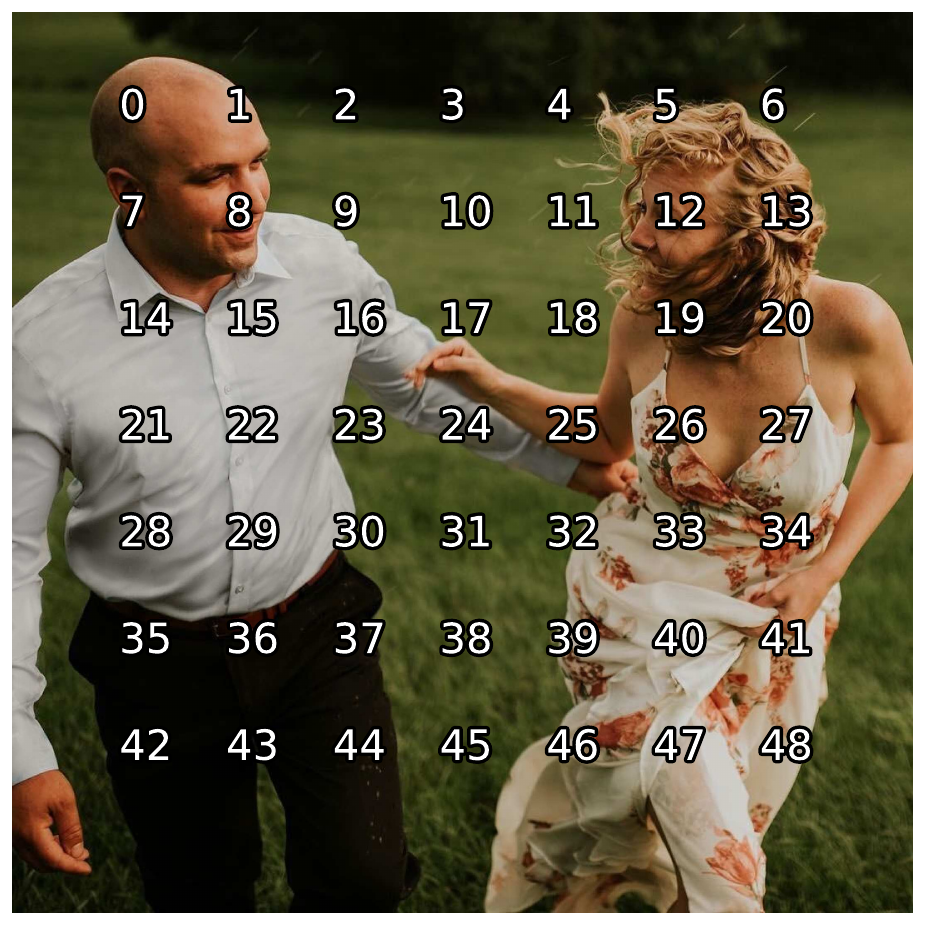}
\caption{Edited Image, with\\ Patch Indices Overlaid}
\end{subfigure}
\hfill\null\\

\null\hfill
\begin{subfigure}[t]{0.9\linewidth}
\centering
\includegraphics[width=\linewidth]{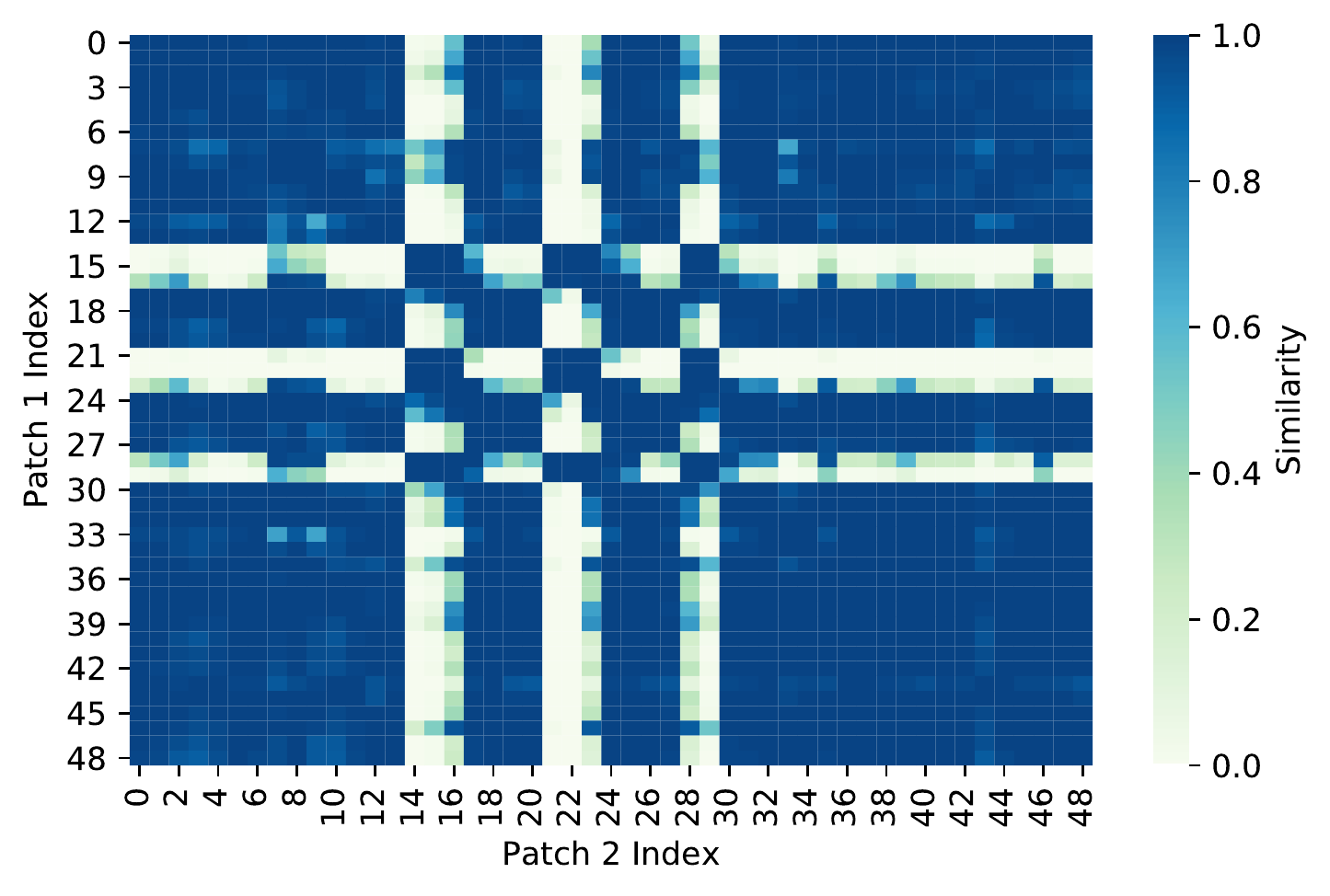}
\caption{Forensic Similarity Matrix}
\end{subfigure}
\hfill\null\\

\null\hfill
\begin{subfigure}[t]{\linewidth}
\centering
\includegraphics[width=\linewidth]{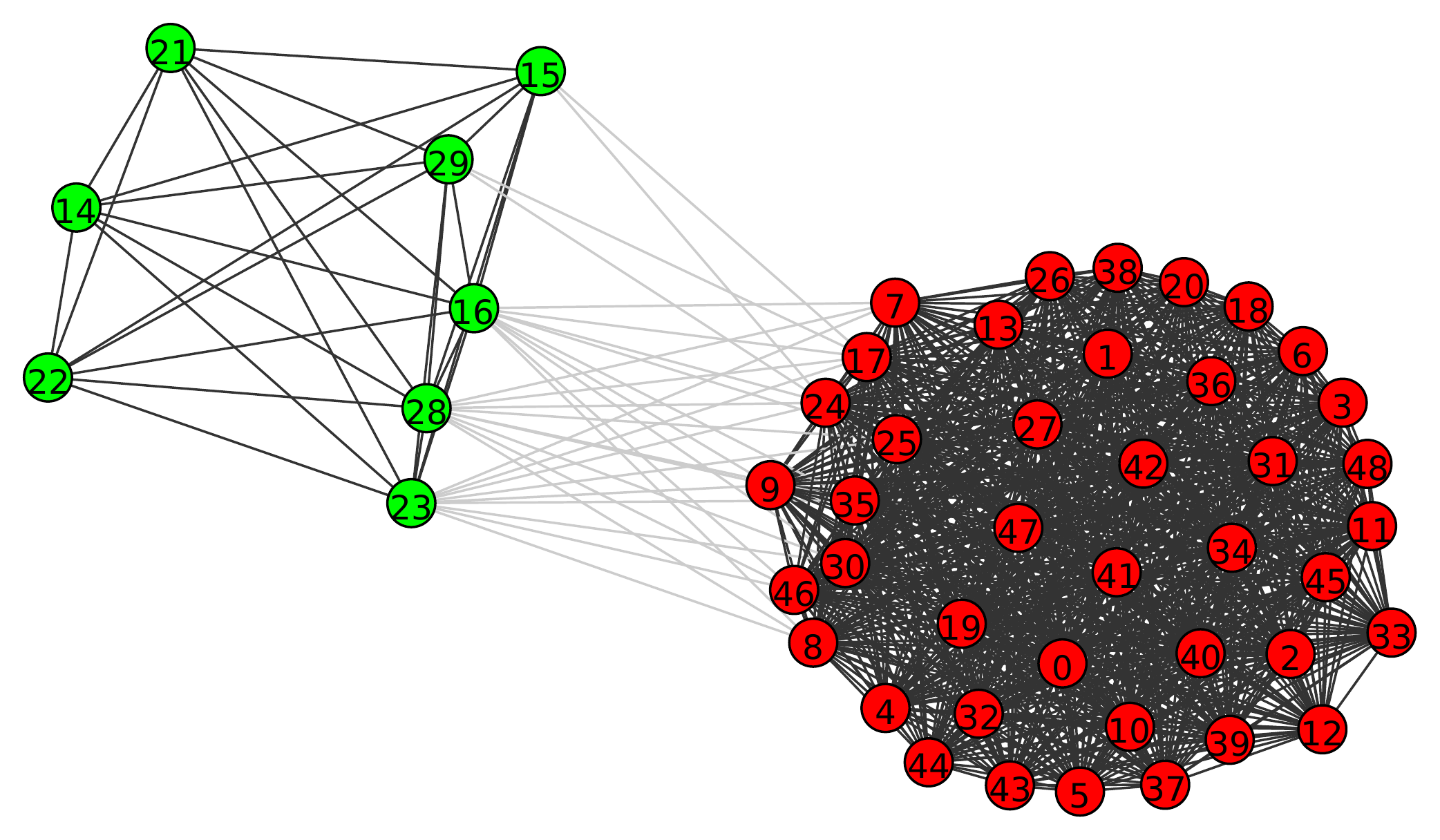}
\caption{Forensic Similarity Graph with Community Partitions}
\end{subfigure}
\hfill\null
\caption{An example of community structure in the forensic similarity graph representation of a forged image. Original image (a) and edited image (b), where water stains were removed using a brush tool, with vertex indices overlaid. Editing credit to Reddit.com user ``/u/rombouts." The forensic similarity matrix in (c) shows the forensic similarity between each sampled image patch. The graph representation in (d) highlights the community structure, showing the patches associated with the tampered region appears as its own in-connected cluster. The graph in (d) is drawn in \textit{igraph}~\cite{igraph}, showing only edges $>0.90$ and using the Kamada-Kawai force-directed layout method~\cite{kamada1989algorithm}.}\label{fig:simgraph_example}
\end{figure}


\section{The Forensic Similarity Graph}
\label{sec:approach}
In this paper, we propose a new graph-based concept called the ``Forensic Similarity Graph" of an image. The Forensic Similarity Graph captures important relationships among small regions of an image. These relationships bear evidence of localized tampering (e.g. splicing). Through analysis of this representation, we detect and localize falsified images. In later sections, we experimentally show that by using this proposed technique, we detect forgeries with higher accuracy than with techniques that do not capture such relationships.

In the Forensic Similarity Graph, each image patch is represented by a vertex. Edges are assigned between each pair of vertices with weight specified by the forensic similarity between those two corresponding image patches. In forged images, tampered regions form communities, where edges within the tampered region are densely connected with high weight, and are disconnected (or at least have low weight) to the regions outside of the tampered region. The term ``community" aligns with literature on graph-based community detection~\cite{fortunato2010community,newman2004finding,girvan2002community}, which are used to analyze community structures in other domains, such as social networks~\cite{yang2013community} and biological networks~\cite{sah2014exploring}. 


The procedure we propose is as follows:
\begin{enumerate}
\item Sample $n$ patches from the image
\item Calculate forensic similarity between all pairs of sampled patches, according to~\cite{mayer2019similarity}
\item Convert the image into its graph representation, called the Forensic Similarity Graph, with patches as vertices and edges assigned according to the forensic similarity between patches
\item Perform forgery detection and/or localization by applying community detection to the Forensic Similarity Graph
\end{enumerate}



%
%
Here, we more formally describe the construction of the Forensic Similarity Graph. To construct the Forensic Graph of an image, we first sample $n$ patches from the image $\left\lbrace X_1, X_2, \ldots, X_n \right\rbrace$, where $X_i \in \mathbb{X}$ is the $i_{th}$ sampled patch and $\mathbb{X}$ is the space of image patches. In this paper we use square patches that are regularly spaced and overlapped, typically we use a patch sizes of $128\times128$ or $256\times256$ and 50\% to 75\% overlap in both x and y directions.
\footnote{
We note that the proposed representation does not strictly require square input patches sampled at regular intervals, just that accurate forensic similarity measurements are made among sufficiently-many small regions of the image. It may be possible to utilize an appropriate similarity measure that operates on non-square patches. Furthermore, smaller and larger amounts of overlap may be utilized, and even non-uniformly spaced or adaptive sampling locations. All that is required is a measure that accurately captures forensic similarity, and that sufficiently many patches are sampled in the tampered region (if it exists) and in the unaltered region.}

We define the graph $G=(V,E)$ with vertex set \mbox{$V = \left\lbrace X_1, \ldots, X_n \right\rbrace$.} 
and edges $E$ which connect each unique pairing of vertices. Edges have weight $W_{i,j}$ equal to the forensic similarity between $X_i$ and $X_j$, given that the similarity is larger than a threshold, and otherwise have weight $0$. That is, the graph has edge weights
%
\begin{align}
W_{i,j} = \begin{cases}
\fsim{X_i}{X_j}, &\fsim{X_i}{X_j} \geq t\\
0, &\fsim{X_i}{X_j} < t\end{cases}
\label{eq:edge_weight_matrix}
\end{align}
where $W_{i,j}$ is the edge weight between vertex $i$ and vertex $j$, and $\fsim{X_i}{X_j} \in [0,1]$ is the forensic similarity between image patch $X_i$ and image patch $X_j$. Unless otherwise specified, we use a threshold $t=0$ meaning that the graph is fully connected, with $\frac{n^2 - n}{2}$ non-zero edges. In some cases, we use a threshold $1 > t > 0$ to improve the visual interpretation of the graph, or to improve algorithm performance.
Here, we use forensic similarity $S(\cdot,\cdot)$ as defined in~\cite{mayer2019similarity}, which quantifies the similarity of the forensic traces across two image patches related to the source and/or processing history. A value of 1 indicates high similarity, and a value of 0 indicates dissimilarity. 
The edge weights create the edge weight matrix, which we refer to as the \textit{Forensic Similarity Matrix}. A visual example of this Forensic Similarity Matrix can be seen in Fig.~\ref{fig:simgraph_example}c. In some literature, an edge weight matrix is sometimes referred to as the weighted adjacency matrix~\cite{von2007tutorial,fortunato2010community}.

Community structures in this Forensic Similarity Graph are then used to identify and localize image forgeries. For example, in an image that has been locally tampered (e.g. a spliced image), an edge between a patch in the tampered region and a patch in the unaltered image region will have no-or-low weight, since they have dissimilar forensic traces. However, edges among patches within a tampered region are densely connected with high weight, provided that the entirety of the tampered region contain the same forensic traces, i.e. undergone the same tampering process (e.g. brush tool, resizing, etc). Additionally, edges among patches in the unaltered part of the image are similarly densely connected with high weight, since all unaltered patches have the same processing history. As a result, this forms subsets of vertices have have high edge-connectivity within the subset, called a community, and low edge-connectivity to other subsets.


In unaltered images, where every patch was captured by the same camera model and has the same processing history, all edges in the graph are expected to have high weight forming only a single community. In addition, images that have been uniformly/globally tampered will also have all edges of high weight, since all patches will have the same processing history. Examples of uniform tampering include global image resizing, JPEG re-compression, and global non-adaptive contrast enhancement. 

Detecting community structures in general graphs has been studied in other domains~\cite{fortunato2010community,girvan2002community,newman2004finding,sah2014exploring}.
In Sec.~\ref{sec:approach}, we adapt two community detection techniques from literature, and apply them to the Forensic Similarity Graph to perform forgery detection and localization. Later, in Sec.~\ref{sec:detection_experiments} and \ref{sec:localization_experiments}, we experimentally show that our proposed community detection based approaches achieve higher forgery detection accuracy than approaches that do not consider community structure.




\subsubsection{Example} We show an example of the graph representation in Fig.~\ref{fig:simgraph_example}. 
The tampered image is sampled by gridding it into $256\times256$ patches, with 50\% overlap. An index is assigned to each vertex and are shown in Fig.~\ref{fig:simgraph_example}(b) at their corresponding patch locations. 
The edge weight matrix is shown in Fig.~\ref{fig:simgraph_example}(c), where the vertices corresponding to the edited shirt patches, indices = $\lbrace 14, 15, 16, 21, 22, 23, 28, 29 \rbrace$, have low similarity to the rest of the image patches, and high similarity to other edited shirt patches. Similarly, the non-edited patches have low similarity to the edited patches, and high similarity to other non-edited patches. This suggests that in this tampered image example, there are two forensic ``communities." 

In Fig.~\ref{fig:simgraph_example}(d), the vertex-edge representation of this graph is shown. 
This visual representation clearly highlights that two communities exist, one associated with the tampered region, and one associated with the non-tampered region. Vertex colors are assigned according to the partitioning determined by the a ``modularity optimization'' method~\cite{newman2004finding}, with edges between the identified communities grayed out. 

\section{Community Detection}
\label{sec:approach:ssec:detection}
In the previous section, we introduced the Forensic Similarity Graph of an image and provided intuition about how structures in this representation, called ``communities," indicate that the image has been locally tampered. In this section, we present techniques to perform the detection of this community structure, and as a result detect and localize the image tampering.

Forgery detection and forgery localization are two related, but very different problems in multimedia forensics, and are differentiated in literature~\cite{bondi2017cvprw,huh2018forensics}. Forgery detection techniques are used to indicate whether tampering has occurred in the image, and output a binary yes/no decision about the image. In this work, we propose community detection techniques that input a forensic graph and output a forgery detection decision,
\begin{align}
\text{Forgery Detection}(G) \rightarrow  & \left\lbrace \text{Image is Unaltered,} \right. \\ & ~ \left. \text{Image is Forged} \right\rbrace,
\end{align} 
We also propose community detection techniques that input a forensic graph, and output a tampering classification for each patch of the image, which is then used to segment the tampered regions of the image,
\begin{equation}
\text{Forgery Localization}(G) \rightarrow \left\lbrace c_1, c_2, \ldots, c_n \right\rbrace,
\end{equation}
where $c_i \in \left\lbrace 1, \ldots, k \right\rbrace$ is the community membership for the $i^{th}$ vertex in the graph, i.e. sampled image patch, and $n$ is the number of sampled image patches. We note that the nature of this type of localization analysis partitions the tampered region from the unaltered, but does not inherently identify which of the partitioned regions is the tampered one or unaltered one. This is why each vertex/patch is mapped into a community identifier $\left\lbrace 1, \ldots, k \right\rbrace$ as opposed to $\left\lbrace\text{Unaltered}, \text{Forged}\right\rbrace$. In this paper we focus on the case of $k=2$, since most evaluation databases only differentiate between the tampered and unaltered regions, and don't differentiate between different tampered regions, if they exist. However, the techniques we propose are general to $k>2$. Recent work has highlighted the need for techniques that identify more than one tampered region~\cite{hosseini2019unsupervised}.

In the remainder of this section, we present two different community detection techniques, namely the Spectral Clustering and Modularity Optimization techniques introduced in Sec.~\ref{sec:background:ssec:community}. These techniques are applied to the Forensic Similarity Graph to detect and partition the community structures associated with image tampering. While we present multiple community detection methods, only one is used at a time, and we compare their efficacy in the experimental results. 
%
Additionally, we show how to convert the patch-based localization partitions into a pixel-level segmentation of forged and unaltered regions. This is done so that the proposed localization methods can be effectively compared to other forgery localization techniques that operate at the pixel-level, such as those in~\cite{cozzolino2015splicebuster,bondi2017cvprw,huh2018forensics,cozzolino2018noiseprint}.

\subsection{Spectral Clustering}
In this method, we apply Spectral Clustering to the forensic similarity graph to detect forgeries and localize tampered regions. We start by constructing the Forensic Similarity Graph an image, described in Sec.~\ref{sec:approach}, by sampling $n$ patches and computing the edge weight matrix $W$. Then, we construct the graph Laplacian $L$ according to \eqref{eq:graph_laplacian}. Forgery detection and forgery localization are then performed as follows. 

\begin{figure}[t]
\null\hfill
\begin{subfigure}[t]{0.29\linewidth}
\centering
\includegraphics[width=\linewidth]{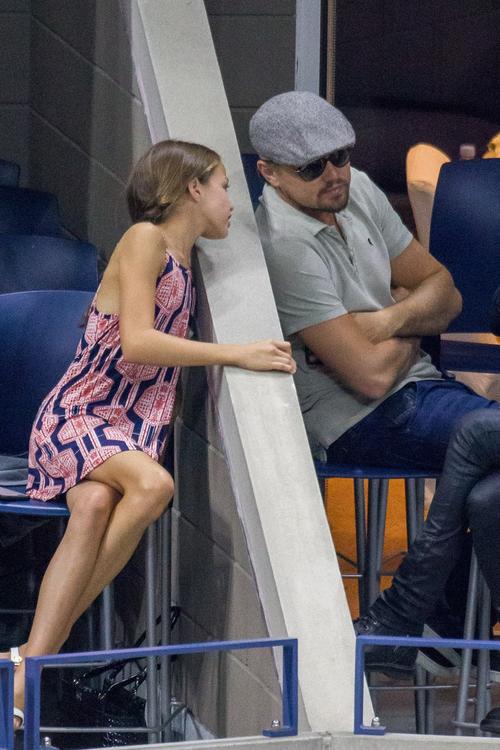}
$\lambda_2=318.42$\\
$Q_{opt}=0.0030$
\caption{Original}
\end{subfigure}
\hfill
\begin{subfigure}[t]{0.29\linewidth}
\centering
\includegraphics[width=\linewidth]{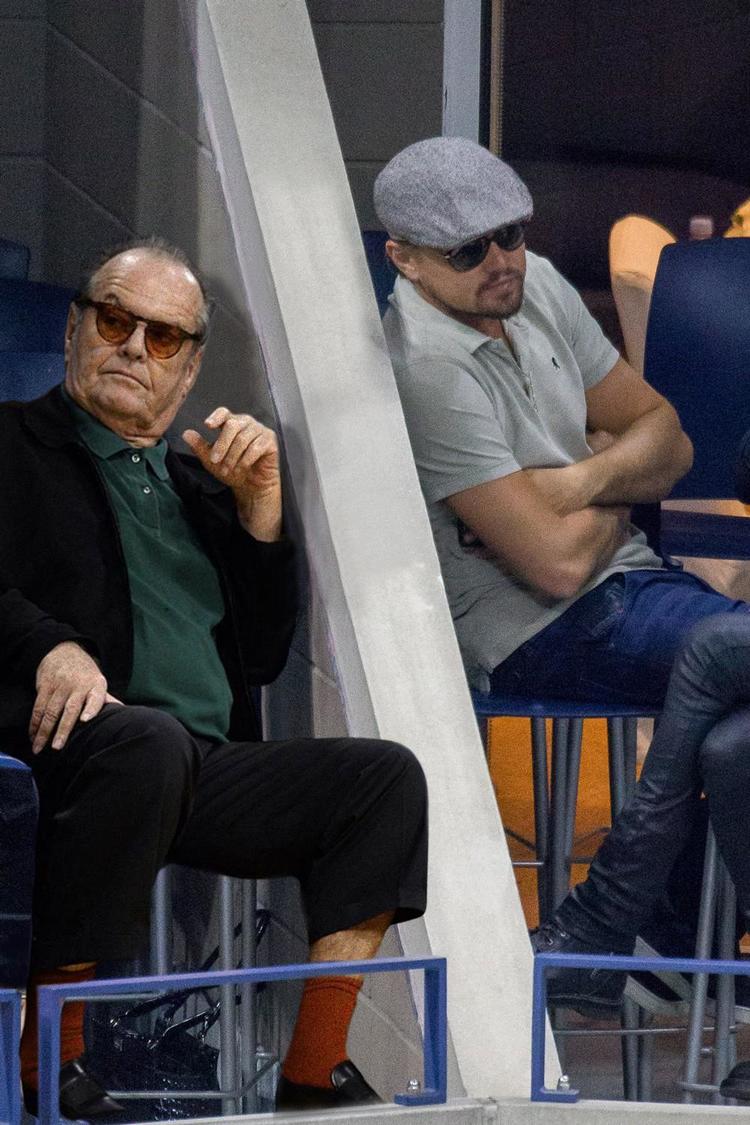}
$\lambda_2=76.75$\\
$Q_{opt}=0.0438$
\caption{Edited}
\end{subfigure}
\hfill\null\\

\null\hfill
\begin{subfigure}[t]{0.32\linewidth}
\centering
\includegraphics[width=\linewidth]{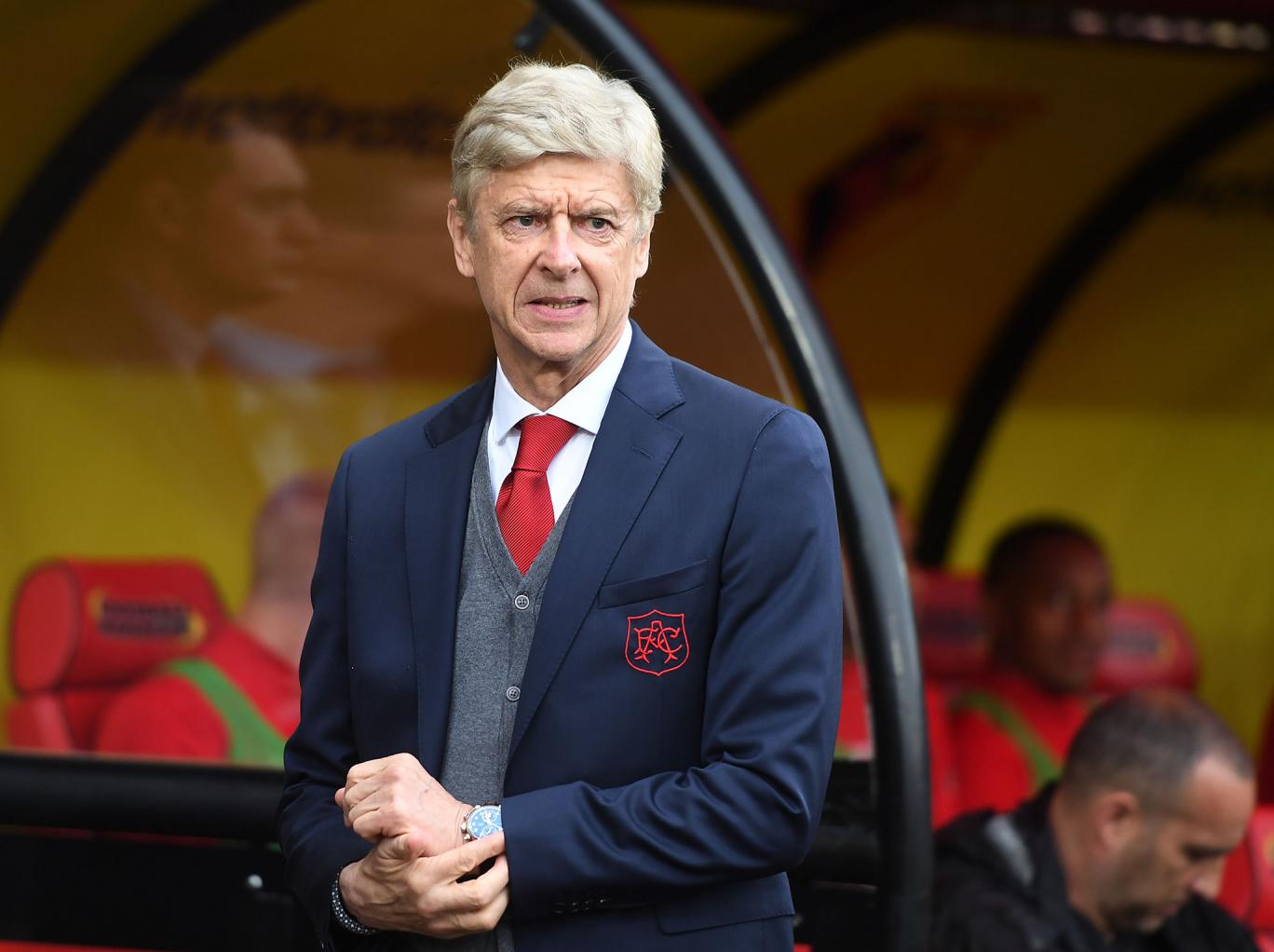}
$\lambda_2=235.56$\\
$Q_{opt}=0.0149$
\caption{Original}
\end{subfigure}
\hfill
\begin{subfigure}[t]{0.32\linewidth}
\centering
\includegraphics[width=\linewidth]{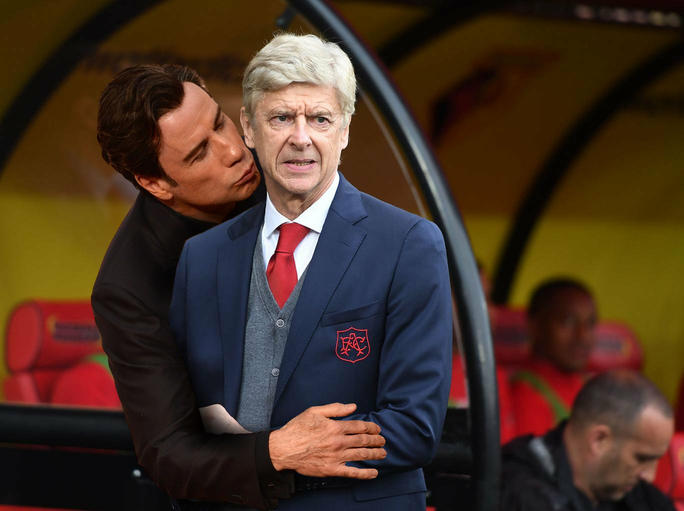}
$\lambda_2=4.28$\\
$Q_{opt}=0.0685$
\caption{Edited}
\end{subfigure}
\hfill\null
\caption{Forgery detection scores on example original and manipulated images. Low $\lambda_2$ values ($<100$) and high $Q_{opt}$ values ($>0.025$) indicate localized tampering.}
\label{fig:detection_examples}
\end{figure}

\subsubsection{Forgery Detection}

As described in Sec.~\ref{sec:related}, the graph Laplacian matrix has a spectrum of eigenvalues $0= \lambda_1 \leq \lambda_2 \leq \ldots \leq \lambda_n$. 
To detect forgeries, we use the fact that the multiplicity of $\lambda=0$ is equal to the number of disconnected communities in the graph to decide whether an image has been forged. Typically, this multiplicity is used to determine the number of communities in graph by determining the number of consecutive eigenvalues less than a threshold, and is sometimes called the ``Eigengap" heuristic~\cite{von2007tutorial}. However, the problem of forgery detection is different in that we need to know only whether the image is forged or unaltered. In other words, we are concerned only if there exists a single community (unaltered) or, alternatively, more than one community (localized tampered).

To do this, in our approach we calculate the second smallest eigenvalue $\lambda_2$. If this value is low, it is indication that at least two communities exist, and potentially more. If it is high, then it is an indication only one community exists, and the image has not been altered. 
As a result, our forgery detection decision rule becomes
\begin{align}
\text{Spectral Gap}\left(G \right) =  \begin{cases}
\text{Unaltered}, & \lambda_2 \geq \tau\\
\text{Forged}, & \lambda_2 < \tau,
\end{cases}
\label{eq:spec_detection}
\end{align}
where $G$ is the forensic similarity graph of the image, $\lambda_2$ is second smallest eigenvalue of the graph Laplacian from Eq.~\eqref{eq:graph_laplacian}, and $\tau$ is the decision threshold determined empirically depending on the desired operating point. This ``spectral gap'' value is also sometimes referred to as the Fiedler eigenvalue, which measures the algebraic connectivity of the graph~\cite{fortunato2010community,fiedler1973algebraic}.

Examples of the spectral gap are shown in Fig.~\ref{fig:detection_examples} for unaltered and forged images. In the two unaltered images(a) and (c) the spectral gap values of 318.42 and 235.56 are high, indicating that only one forensic community exists and no tampering has occurred. In the two spliced images (b) and (d) the spectral gap values of 76.75 and 4.28 are low, indicating that more than one forensic community exists and the tampering has occurred.

\subsubsection{Forgery Localization}
To perform localization, we use the fact that the eigenspace of $0$ of the graph Laplacian is spanned by the indicator vectors $\mathbbm{1}_{A_i}$ to partition the graph vertices into forged and unaltered regions. In this paper we consider the case of $k=2$, in order to partition the tampered versus unaltered regions. 
To do this, we first calculate the eigenvector $\mathbf{u}_2 \in \mathbb{R}^n$ associated with the eigenvector $\lambda_2$
\begin{equation}
\mathbf{L} \mathbf{u}_2 = \lambda_2 \mathbf{u}_2,
\end{equation}
where $\lambda_2$ is the second smallest eigenvalue of the graph Laplacian matrix.

Next, we use the sign of each component of $\mathbf{u}_2$ to assign each vertex into a community membership,
\begin{align}
\hat{c}_i =  \begin{cases}
1, & u_{i,2} \geq 0\\
2, & u_{i,2} < 0,
\end{cases}
\end{align}
where $\hat{c}_i$ is the predicted community membership of the $i^{\text{th}}$ graph vertex / image patch. This method approximates the (NP-Hard) Ratio cut algorithm for $k=2$. When using the normalized graph Laplacian, it approximates the Normalized Cut algorithm~\cite{von2007tutorial}.

\begin{figure}[t]
\centering
\includegraphics[width=0.66\linewidth]{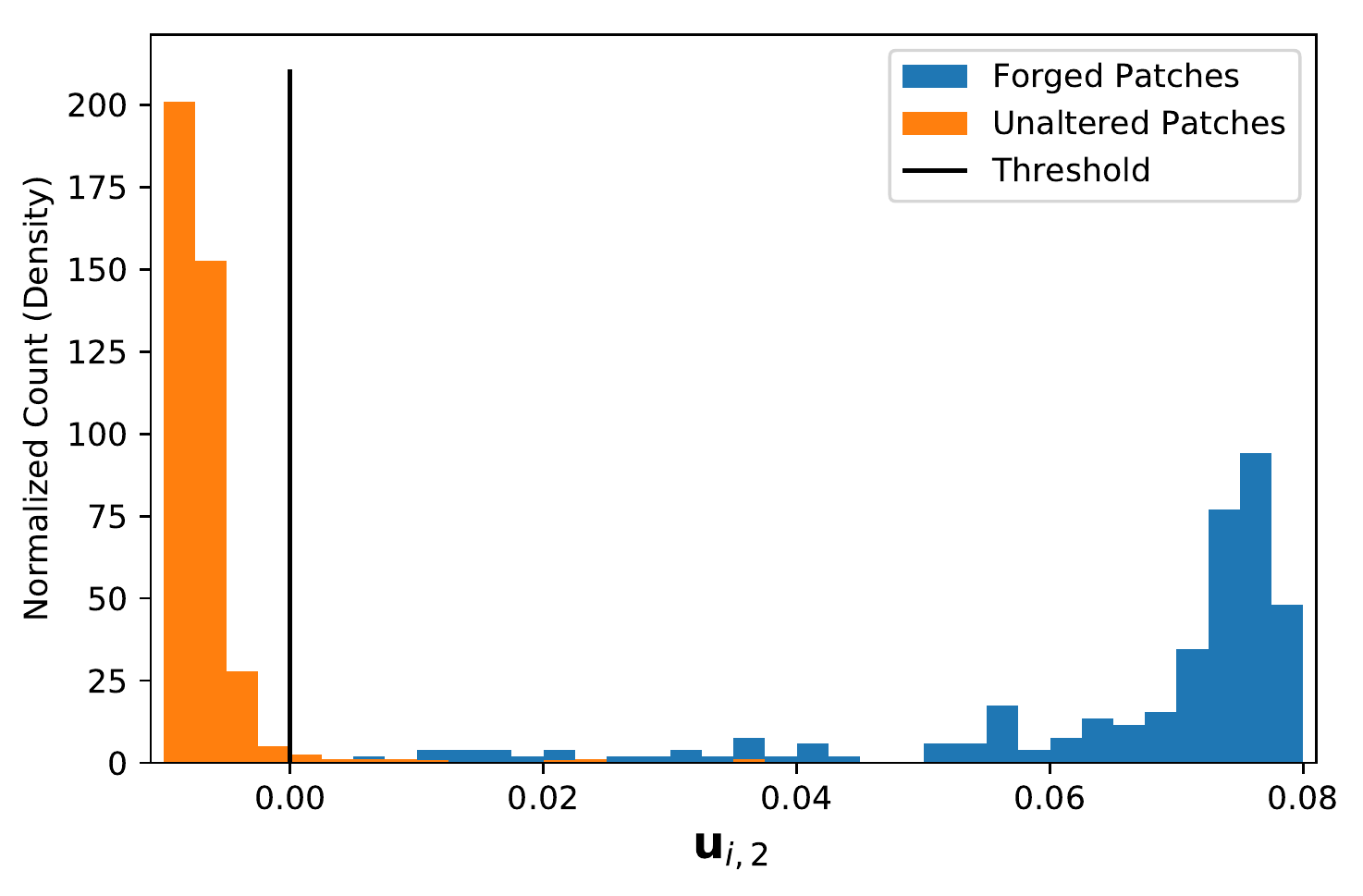}
\caption{Component values of $\mathbf{u}_2$ for the edited image in Fig.~\ref{fig:detection_examples}(d). In this example, components of $\mathbf{u}_2$ corresponding to the forged patches are greater than 0, and unaltered patches are less than 0.}
\label{fig:sc_partition_example}
\end{figure}

An example of this algorithm is shown in Fig.~\ref{fig:sc_partition_example}, where we partition the image patches from the spliced image in Fig.~\ref{fig:detection_examples}(d) into two communities. In this figure, we show the histogram of values of $\mathbf{u}_2$. The components associated with patches in the tampered region are shown in blue, and are typically above 0. The components associated with patches in the unaltered regions are shown in orange, and are typically below 0. Ground truth was determined based on the central location of each image patch.

We note that this algorithm can be easily extended to consider $k>2$. This is done by considering the eigenvector values for each vertex as a point in space and performing k-means clustering~\cite{von2007tutorial}. 

\subsection{Modularity Optimization}
Here, we describe a second technique for forgery detection and localization based on Modularity Optimization. We start by constructing the Forensic Similarity Graph an image, described in Sec.~\ref{sec:approach}, by sampling $n$ patches and computing the edge weight matrix $W$. We then perform the Modularity optimization~\eqref{eq:mod_optimization} resulting in $Q_opt$.

To calculate, $Q_{opt}$ we use a popular modularity optimization technique called the ``fast-greedy'' method, which agglomeratively forms communities by iteratively combining communities which yield the greatest modularity increase~\cite{newman2004fast,clauset2004finding}, and reports the largest modularity value found. While there are a number of variations of modularity optimization techniques, the fast-greedy method is a popular technique~\cite{fortunato2010community} and operates on a reasonable timescale on our forensic graphs. Furthermore, when constructing $W$, we typically use a thresholded edge weight matrix~\eqref{eq:edge_weight_matrix}, often with a threshold $0.5 < t < 0.9$, for modularity optimization techniques. Using a thresholded edge weight matrix significantly reduces the complexity of the optimization, and we have anecdotally found that doing so produces more reliable modularity scores.

\subsubsection{Forgery Detection}
To detect whether the image has been locally tampered, we compare the optimized modularity value, $Q_{opt}$, found for the optimal partition to a threshold. If $Q_{opt}$ exceeds the threshold, it indicates that there is strong community structure and, as a result, indicates that one or more regions of the image has been tampered with. Our decision measure becomes 
\begin{align}
\text{Modularity Opt}\left(G \right) =  \begin{cases}
\text{Unaltered}, & Q_{opt} < \tau\\
\text{Forged}, & Q_{opt} \geq \tau,
\end{cases}
\label{eq:mod_detection}
\end{align}
where $G$ is the forensic similarity graph of the image, $Q_{opt}$ is the optimized modularity value found on $G$, and $\tau$ is the decision threshold, chosen empirically.

\subsubsection{Forgery Localization}

Here, we localize tampered regions by using the optimized community memberships, $\hat{c}$, to assign each patch into a community,
\begin{equation}
\{\hat{c}_1, \ldots, \hat{c}_n \} = \argmax_{\{c_1, \ldots, c_n \}} \frac{1}{4m}\sum_{i,j}\left(W_{ij}-\frac{d_id_j}{2m}\right)\mathbbm{1}(c_i = c_j)
\end{equation}
where $\hat{c}_i \in \lbrace 1, \ldots, k \rbrace$ is the optimized community membership of the $i^\text{th}$ vertex/image patch determined by the modularity optimization algorithm. As with the Spectral Clustering method, we use $k=2$ to segment the unaltered region from the forged region(s). Though we note that this method is trivially extendable to $k>2$.

As an example, the partitioning into the green and red coloring of Fig.~\ref{fig:simgraph_example} was determined by the Modularity Optimization method, using an edge weighting threshold of $t=0.9$ and $k=2$.\looseness=-1

\subsection{Pixel-level localization}
\label{sec:approach:ssec:pixel_level}
The above localization techniques partition image patches into forged and unaltered regions. However, localization is often performed at the pixel level~\cite{huh2018forensics,bondi2017cvprw,cozzolino2015splicebuster}. 
Here, we describe how to convert patch-level community partitions to a pixel-level forgery localization.

We start by building a pixel map of localization predictions $\mathbf{P}$ with dimensions in x and y that are the same as the image,
\begin{equation}
P_{x,y} =  \sum_{l=1}^N \mathbbm{1}\left[c_l = \alpha \right] \mathbbm{1}\left[(x,y) \in R_l \right],
\end{equation}
where $c_l$ is the predicted community membership for patch $l$, and $R_l$ is the set of coordinates covered by patch $l$. That is, for each pixel we sum the number of patches that contain that pixel and are also predicted to be in community \mbox{$\alpha \in \left\lbrace1, \ldots, k\right\rbrace$}. The investigator can vary $\alpha$ to inspect the various communities. For $k=2$, we choose $\alpha$ to be the smaller of the two communities, since the largest community is likely to be the unaltered region. 

Pixels near the edges and corners of an image are covered by less patches, and are more difficult to accurately predict. To address this, we create a normalized pixel map
\begin{equation}
P^{Norm}_{x,y} = \frac{P_{x,y}}{T_{x,y}},
\end{equation}
where $\mathbf{T}$ is the map of total patches that cover each pixel defined by
\begin{equation}
T_{x,y} = \sum_{l=1}^N\mathbbm{1}\left[(x,y) \in R_l \right].
\end{equation}

Finally, we smooth the normalized pixel map, using a Gaussian blur kernel, and compare to a threshold, chosen empirically.

\begin{figure}[t]
\null\hspace{0.1mm}
\begin{subfigure}[t]{0.4\linewidth}
\centering
\includegraphics[width=\linewidth]{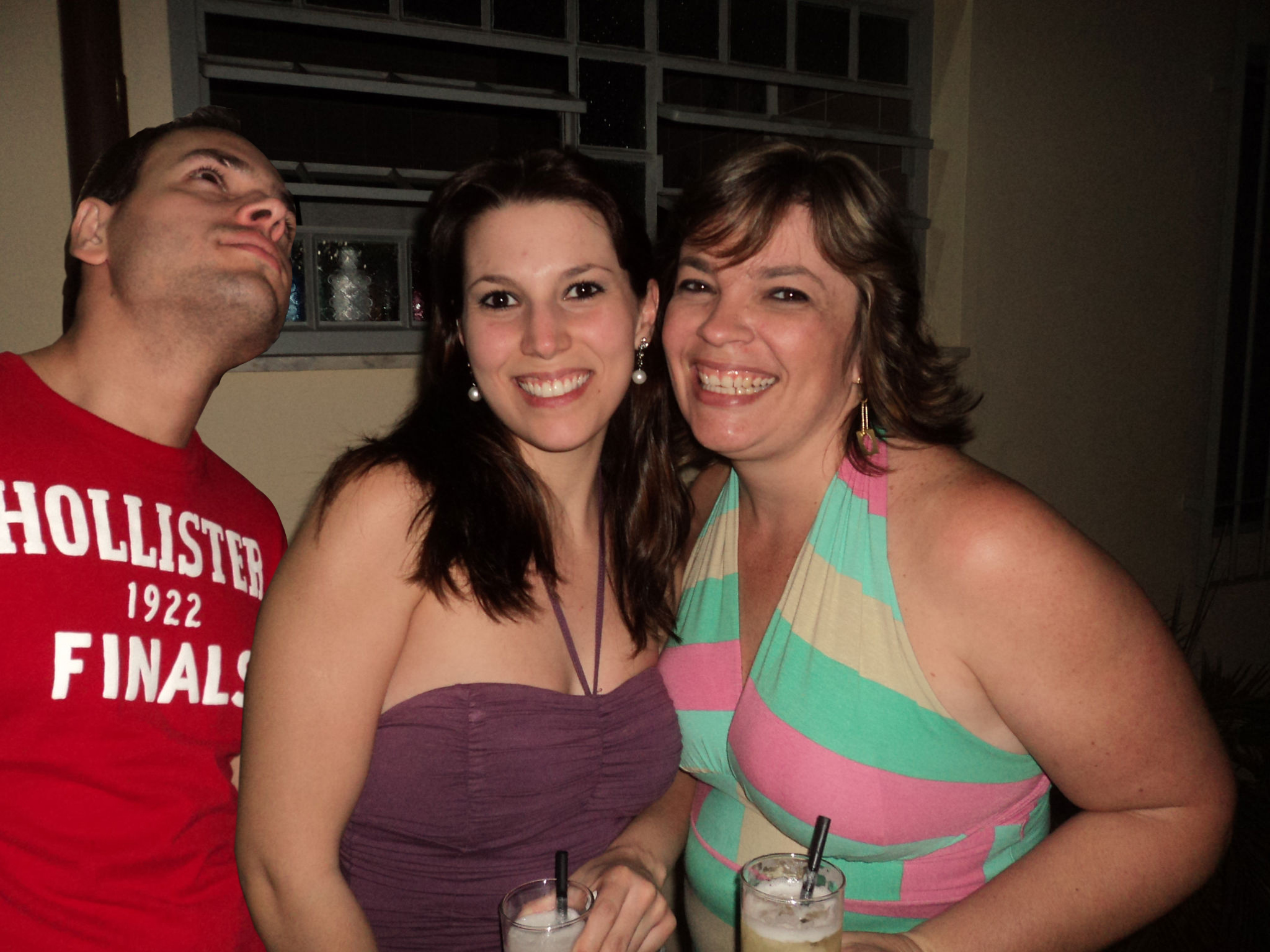}
\caption{Edited Image}
\end{subfigure}
\hfill
\begin{subfigure}[t]{0.4\linewidth}
\centering
\includegraphics[width=\linewidth]{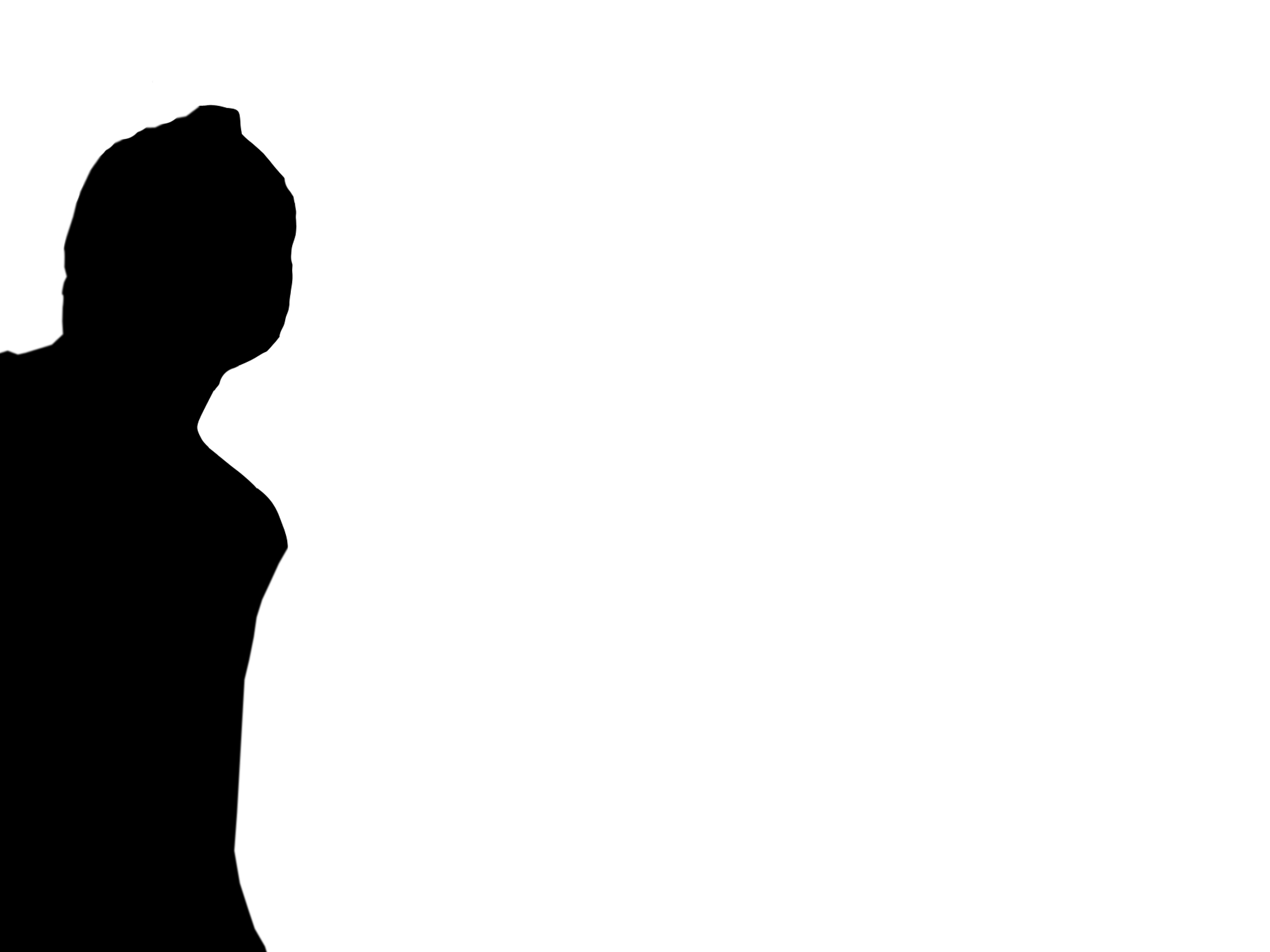}
\caption{Ground Truth}
\end{subfigure}
\hfill\null

\null
\begin{subfigure}[t]{0.49\linewidth}
\centering
\includegraphics[width=\linewidth]{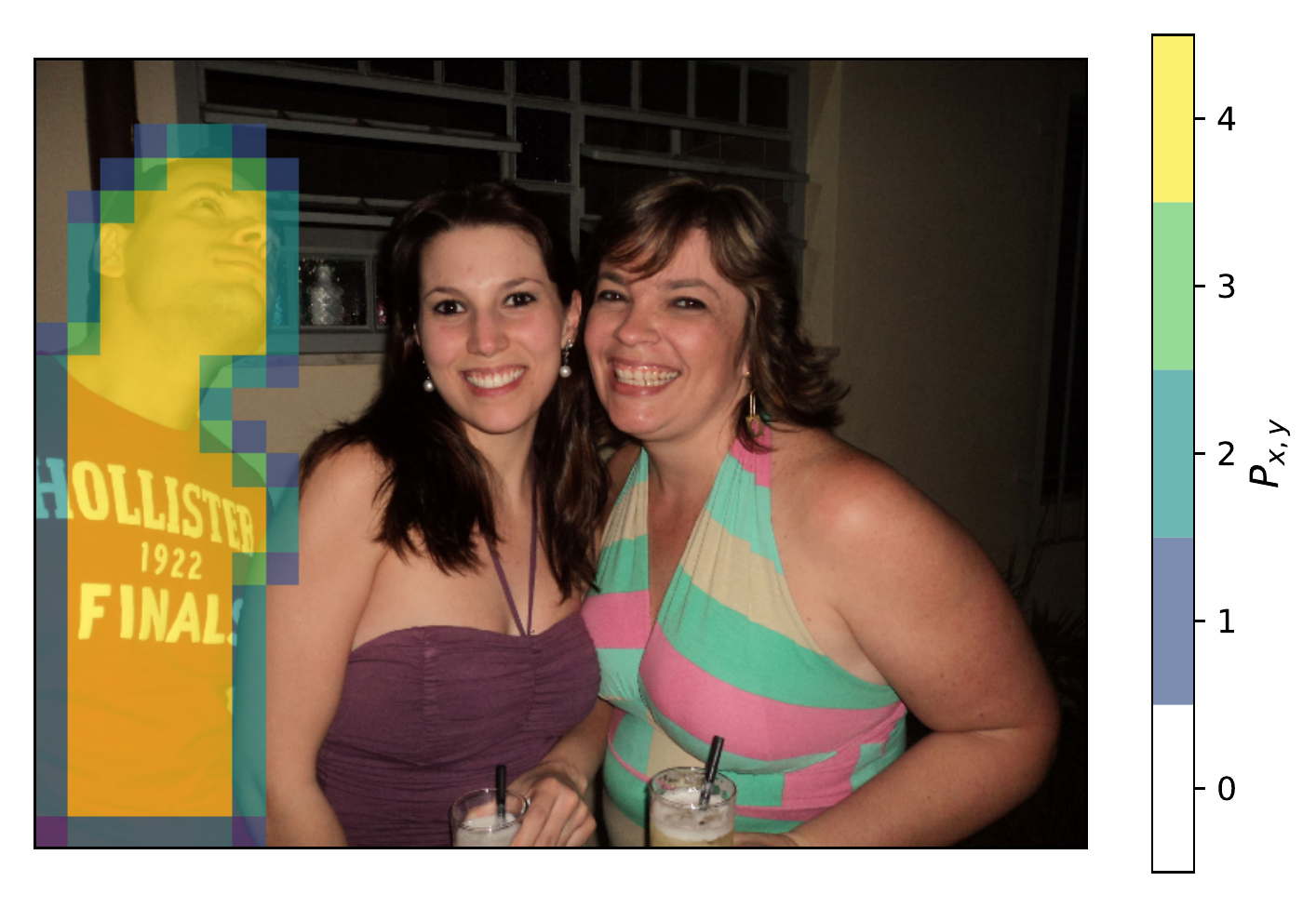}
\caption{Prediction Map}
\end{subfigure}
\hfill
\begin{subfigure}[t]{0.49\linewidth}
\centering
\includegraphics[width=\linewidth]{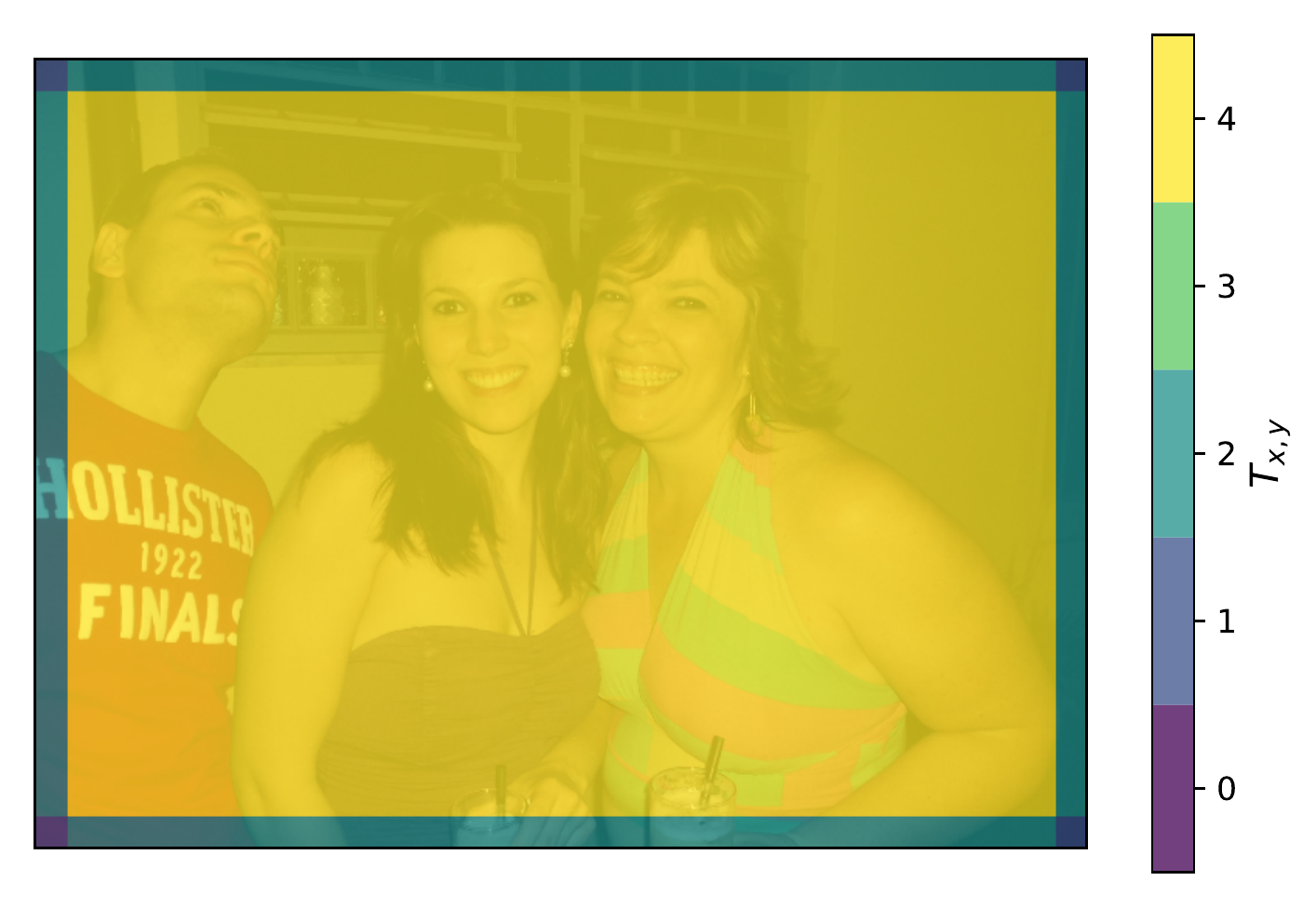}
\caption{Patch Coverage Map}
\end{subfigure}
\hfill\null

\null
\begin{subfigure}[c]{0.49\linewidth}
\centering
\includegraphics[width=\linewidth]{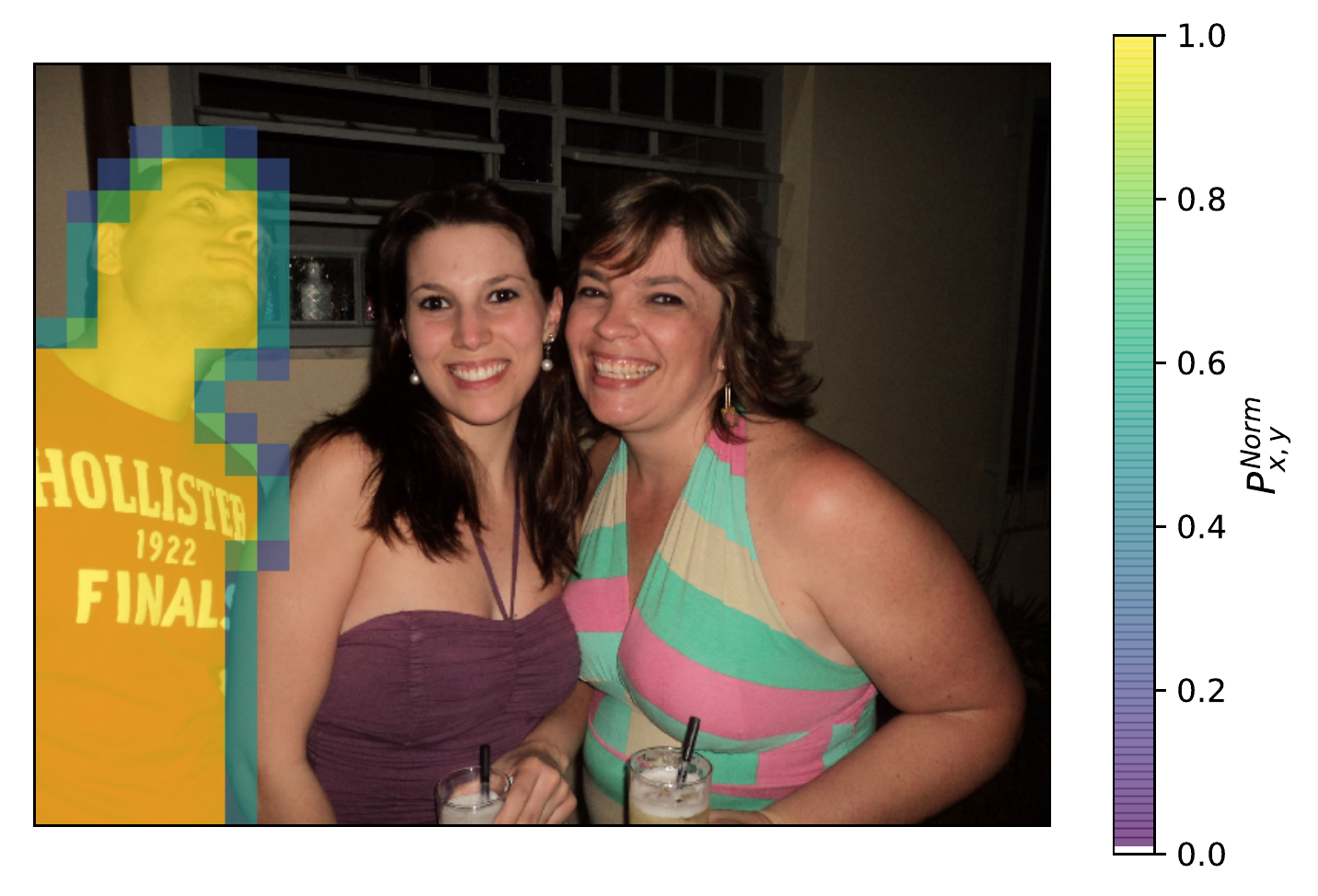}
\caption{Normalized Prediction Map}
\end{subfigure}
\hfill
\begin{subfigure}[c]{0.41\linewidth}
\centering
\includegraphics[width=\linewidth]{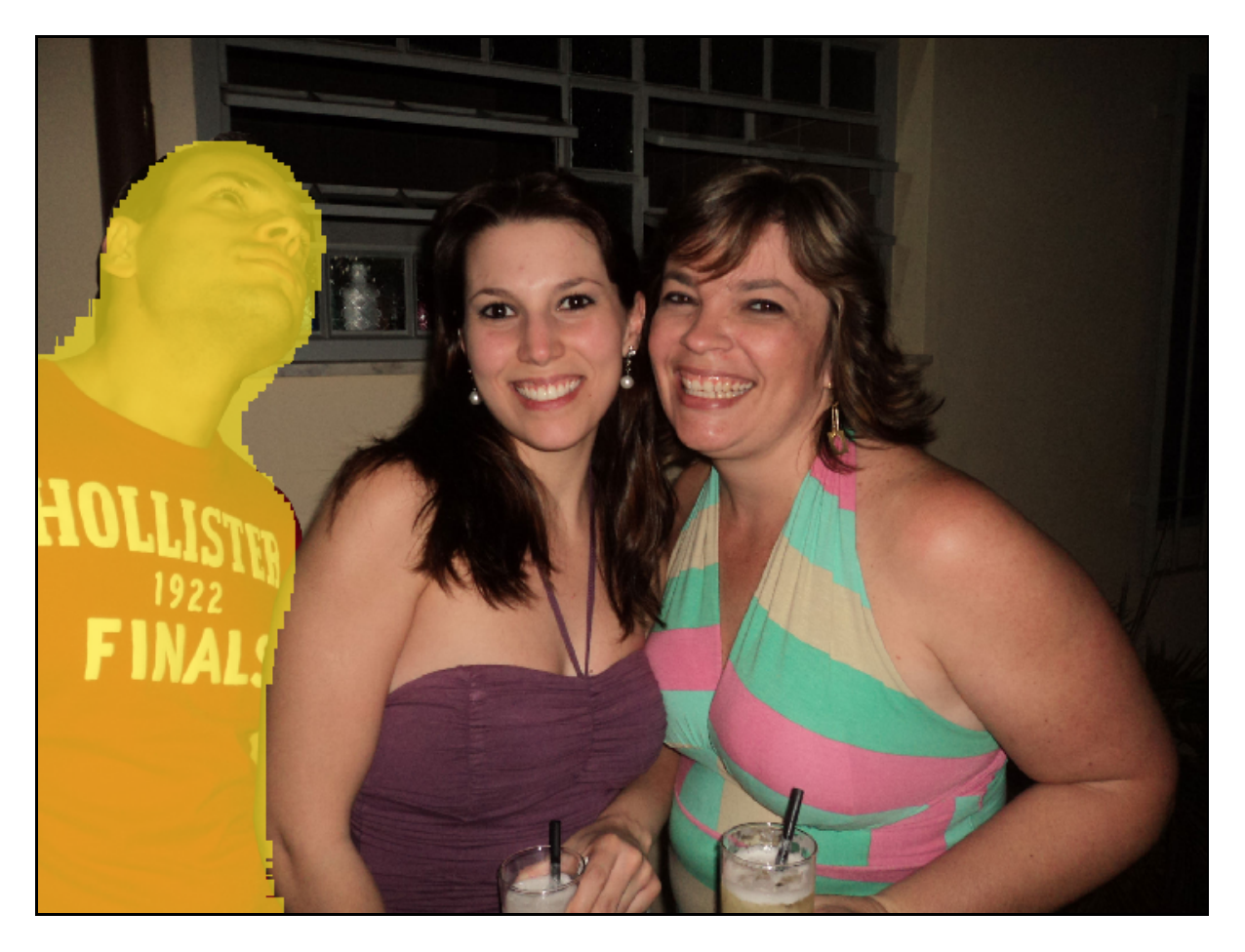}
\caption{Smoothing + Threshold}
\end{subfigure}
\hspace{5.5mm}\null
\caption{An example showing the conversion from patch-level community partitions to pixel-level forgery localization, described in Sec.~\ref{sec:approach:ssec:pixel_level}.}
\label{fig:pixel_localization_example}
\end{figure}

In Fig.~\ref{fig:pixel_localization_example}, we show an example. Fig.~\ref{fig:pixel_localization_example}a shows an edited image from the Carvalho database~\cite{carvalho2013exposing} and Fig.~\ref{fig:pixel_localization_example}b shows the ground truth mask, with the edited region highlighted in black. Fig.~\ref{fig:pixel_localization_example}c shows the patch-level prediction map, determined using the Spectral Clustering method. Since patches were sampled from image with 50\% overlap, pixels are covered by 1 to 4 patches, as shown in Fig.~\ref{fig:pixel_localization_example}d. Fig.~\ref{fig:pixel_localization_example}e shows the normalized prediction map, in which we can see more confident forgery localization along the corner and edges of the image. Finally, Fig.~\ref{fig:pixel_localization_example}f shows the smoothed and thresholded version of the normalized prediction map, in which the tampered region is localized and matches the ground truth.

\section{Experimental Results: Forgery Detection}
\label{sec:detection_experiments}

We conducted a series of experiments to test the efficacy of our proposed approach for forgery detection and localization. In this section, we describe the experimental procedures and results for forgery detection. In Sec.~\ref{sec:localization_experiments}, we conduct experiments to test forgery localization performance. In general, we found that the proposed Spectral Gap technique achieved highest forgery detection performance, and outperformed naive methods that do not include community structure.

We performed two sets of experiments for forgery detection. In the first experiment, we evaluated performance on three publicly available benchmark datasets, which contain unaltered and tampered images. In the second experiment, we evaluated performance on ``synthetic'' forgeries, which were created by copying and pasting a block from one image into another. While not realistic in appearance, the synthetic forgery experiment shows statistical characterizations of the system, controlling for the size of the forged area.

\begin{figure}[t]
\null\hfill
\begin{subfigure}[t]{0.29\linewidth}
\centering
\includegraphics[width=\linewidth]{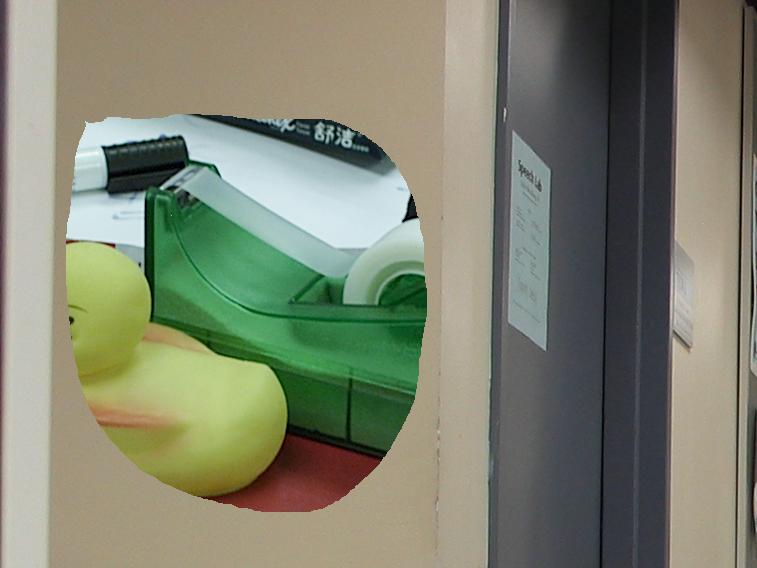}
\caption{Columbia~\cite{hsu06crfcheck}}
\end{subfigure}
\hfill
\begin{subfigure}[t]{0.29\linewidth}
\centering
\includegraphics[width=\linewidth]{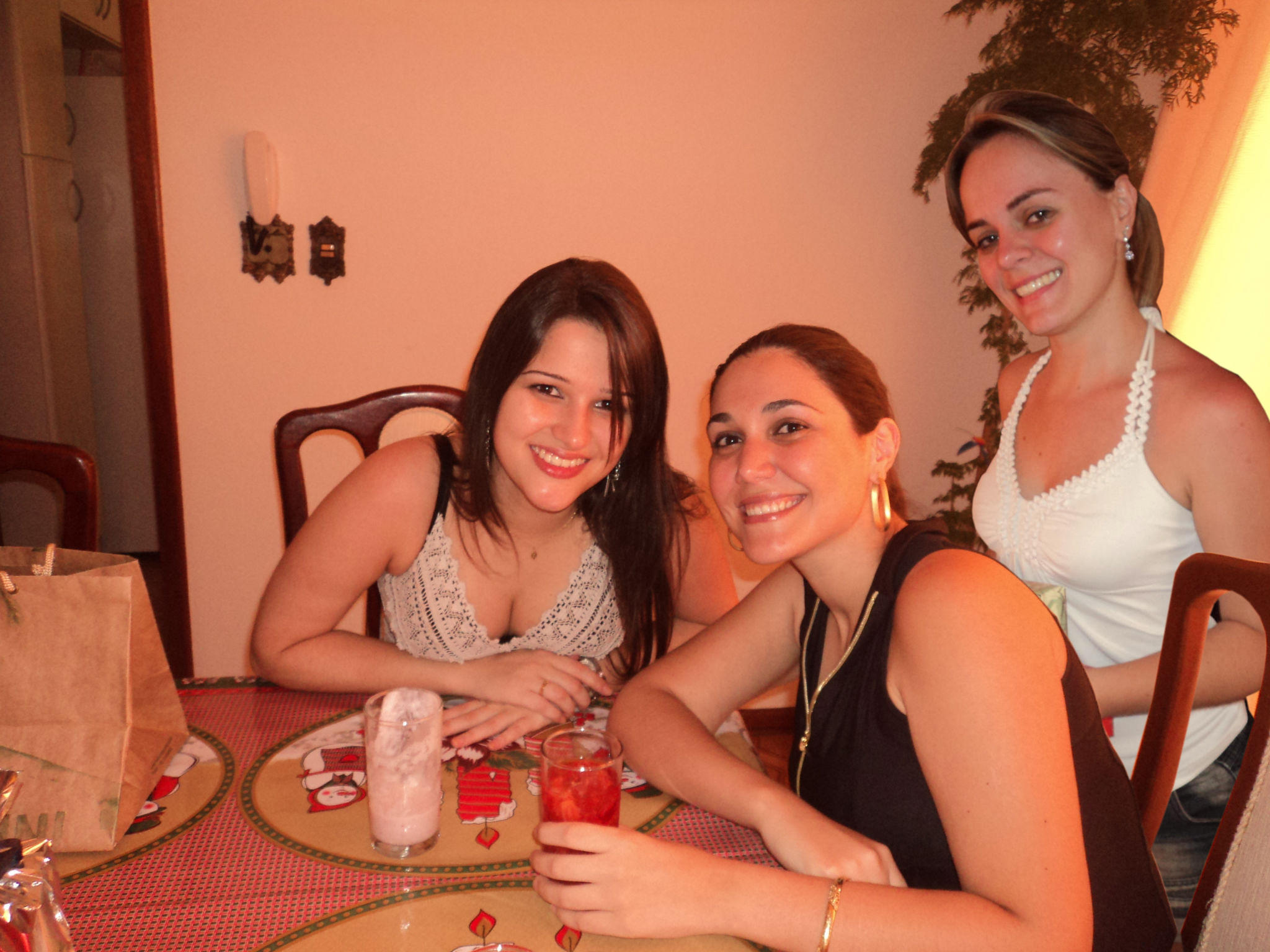}
\caption{Carvalho~\cite{carvalho2013exposing}}
\end{subfigure}
\hfill
\begin{subfigure}[t]{0.385\linewidth}
\centering
\includegraphics[width=\linewidth]{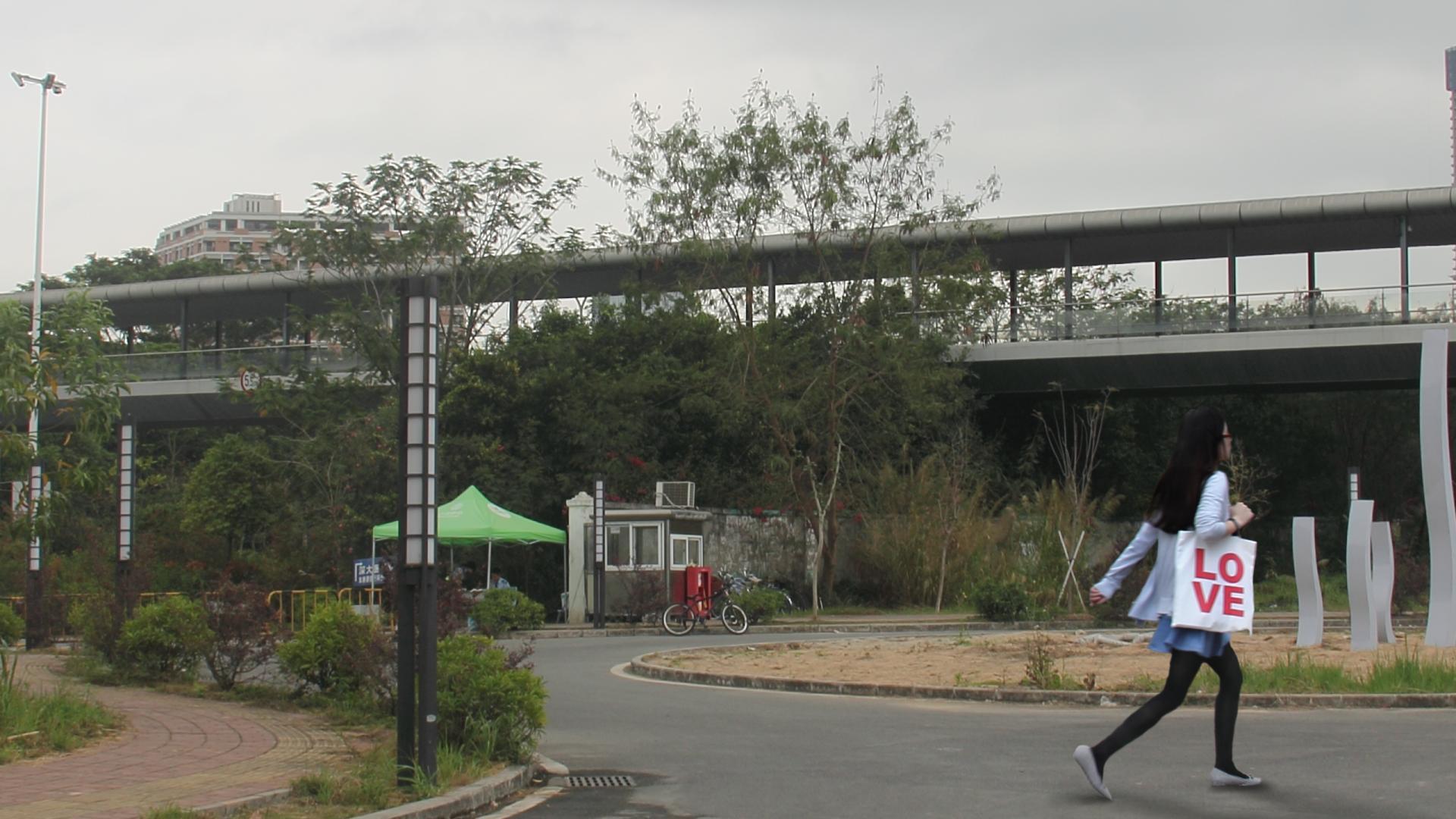}
\caption{Korus~\cite{Korus2017TIFS}}
\end{subfigure}
\hfill\null
\caption{Example forgeries from the benchmark datasets.}
\label{fig:db_examples}
\end{figure}

In these experiments we evaluated the two proposed community detection techniques that are described in Sec.~\ref{sec:approach:ssec:detection}: the Spectral Gap method~\eqref{eq:spec_detection} and Modularity Optimization method~\eqref{eq:mod_detection}. We then compared to two naive approaches that utilize the same forensic similarity procedure of our proposed approach, but do not consider community structure. These are 1) ``Mean Similarity'' which is the average forensic similarity edge weight as a detection measure and used in prior work~\cite{mayer2019similarity}, and 2) ``Minimum Similarity'' using the minimum forensic similarity edge weight as a detection measure. A significant contribution of the Forensic Similarity Graph is that it captures complex relationships among regions of the image, in the form of communities, which enables more powerful analysis. The following experiments show that the proposed community detection based approaches outperform these naive approaches. These results highlight that identifying community structure is critical to accurately detecting image forgeries.

In addition, we compared against several state-of-the-art methods. We compared against the work by Huh et al.~\cite{huh2018forensics}, which computes a deep-learning based self-consistency map and detects image forgeries by spatially averaging this map, and comparing to a threshold. They compute three types of self-consistency maps based on camera-model consistency, image consistency, or EXIF consistency. Our approach is most comparable to the ``Camera'' approach in~\cite{huh2018forensics}, since forensic similarity network we used was trained on patches labeled according to their source camera model.
We also compared to the work by Bondi et al.~\cite{bondi2017cvprw}, in which an 18-elements feature vector is extracted with a CNN trained for camera model identification on 64 x 64 pixels patches. A simple clustering procedure is iteratively applied to separate the background from the foreground.

In these experiments, we calculated the forensic similarity of image patches using the deep-learning approach described in~\cite{mayer2019similarity}, trained on four million image patches from 80 camera models. Two networks were tested, one with forensic patch size 128$\times$128, and a second with size 256$\times$256. Patches were sampled spanning the image with 50\% overlap. The forensic similarity graph was constructed for each testing image as described in Sec.~\ref{sec:approach}. Finally a forgery decision was rendered based on the proposed detection methods in Sec.~\ref{sec:approach:ssec:detection}. Performance was evaluated by the receiver operating characteristic (ROC) performance and the mean-Average-Precision (mAP) score, which were determined by sweeping the decision threshold $\tau$ for each method.

\begin{figure}[t]
\null\hfill
\begin{subfigure}[t]{0.49\linewidth}
\centering
\includegraphics[width=\linewidth]{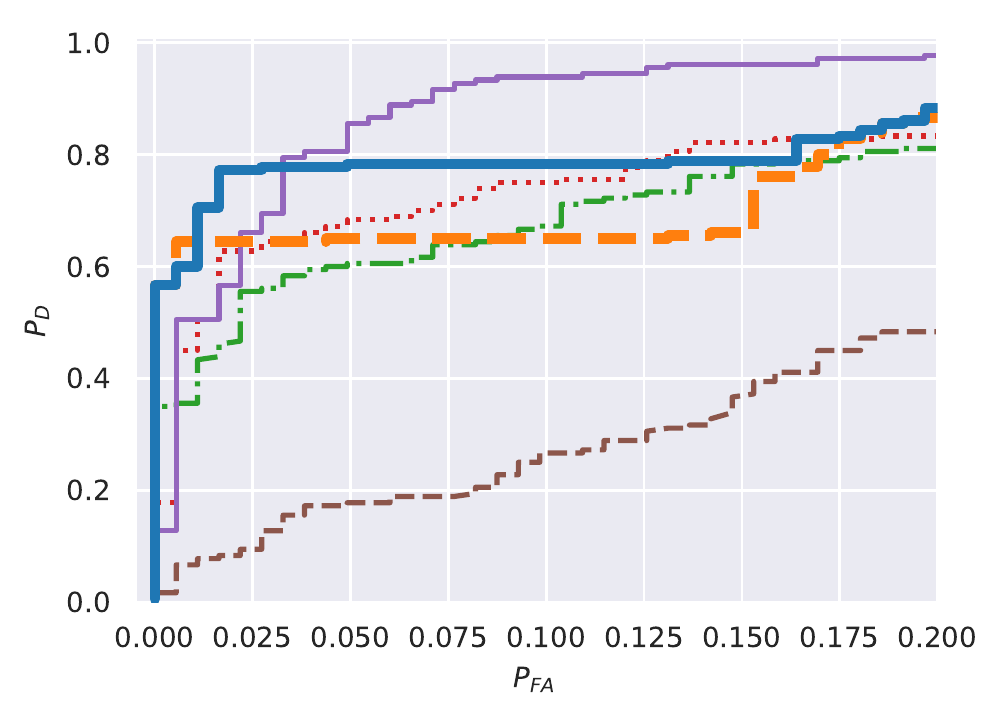}

\vspace{-0.5em}\caption{Columbia (Low PFA)}
\end{subfigure}
\hfill
\begin{subfigure}[t]{0.49\linewidth}
\centering
\includegraphics[width=\linewidth]{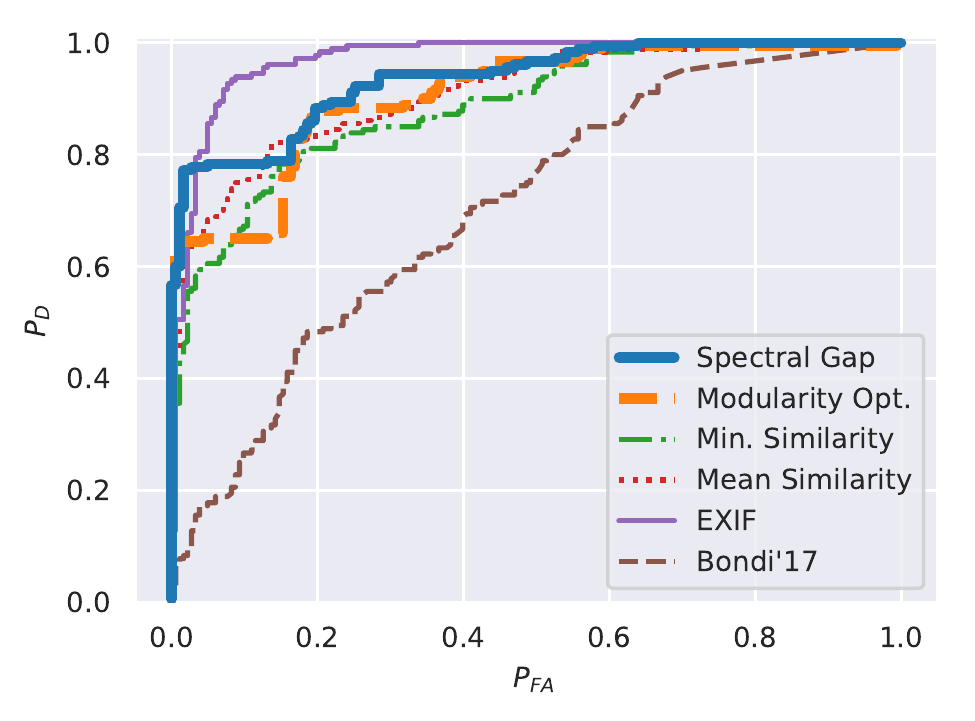}

\vspace{-0.5em}\caption{Columbia}
\end{subfigure}
\hfill\null

\null\hfill
\begin{subfigure}[t]{0.49\linewidth}
\centering
\includegraphics[width=\linewidth]{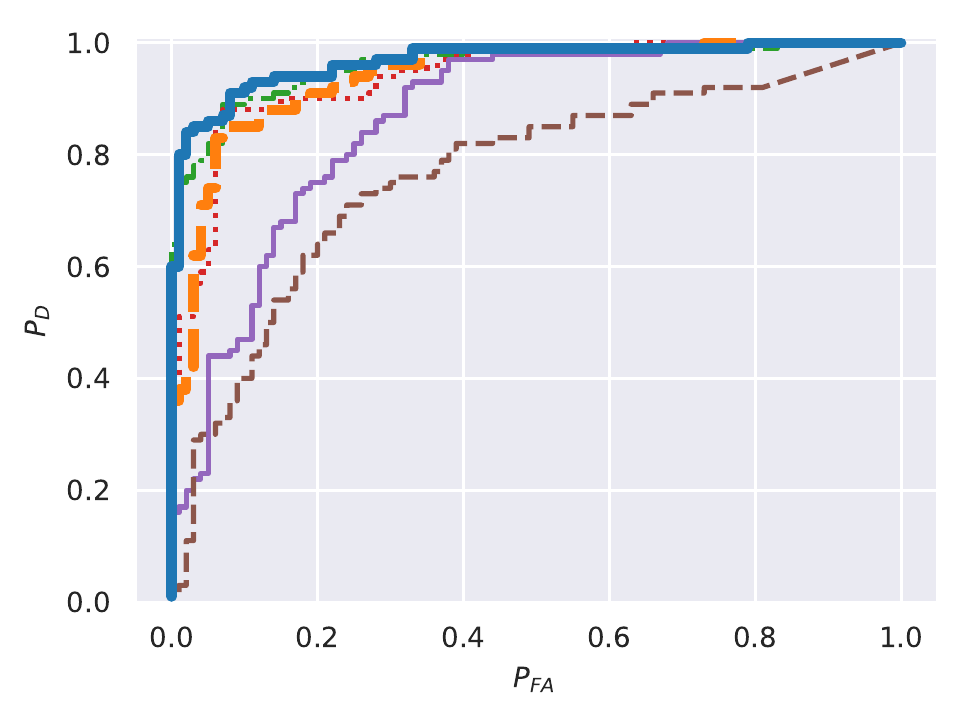}

\vspace{-0.5em}\caption{Carvalho}
\end{subfigure}
\hfill
\begin{subfigure}[t]{0.49\linewidth}
\centering
\includegraphics[width=\linewidth]{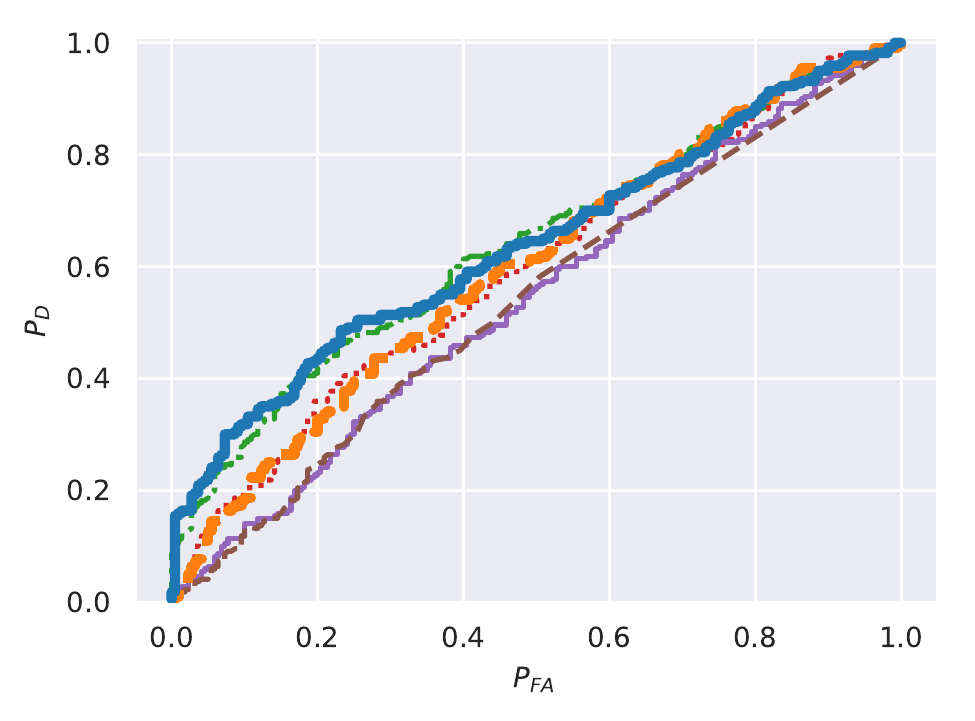}

\vspace{-0.5em}\caption{Korus}
\end{subfigure}
\hfill\null
\caption{Forgery detection ROC curves on benchmark datasets}
\label{fig:db_roc}
\end{figure}

\subsection{Performance on Benchmark Datasets}
In this experiment, we evaluated the performance of our approach on publicly-available benchmarking datasets. These databases are 1) the ``Columbia Uncompressed Splicing Database''~\cite{hsu06crfcheck} containing 180 spliced and 183 authentic tiff images ranging from sizes $757\times 568$ to $1152\times 768$, 2)~the ``Carvalho DSO-1 Database''~\cite{carvalho2013exposing} containing 100 spliced and 100 authentic png images of size $2048\times 1536$, and 3)~the ``Korus Realistic Tampering Dataset''~\cite{Korus2017TIFS} containing 220 tampered and 220 corresponding original tiff images of size $1920\times 1080$. Examples of spliced images from each database are shown in Fig.\ref{fig:db_examples}. 

Fig.~\ref{fig:db_roc} shows ROC curves for our proposed approaches, as well as the methods we compare against. The ROC curve shows the trade off of the probability of correct detection ($P_D$) of a forged image versus the probability of false alarm ($P_{FA}$), i.e. misclassification, of an unaltered image.  In the majority of cases, our proposed approaches achieved highest performance, with the Spectral Gap method achieving highest detection rates. We note that the Huh et al. EXIF method achieves higher detection rates at high false alarm rates in the Columbia dataset, but was outperformed by our approaches at low false alarm rates. 
This is important because in many realistic conditions, forensic investigators must operate at low false alarm rates. This is because scenarios such as criminal or intelligence investigations, there can be a high cost to misidentifying an unaltered image as falsified.

\renewcommand{\arraystretch}{0.85}
\begin{table}[t]\centering
\caption{Mean Average Precision on Benchmark Databases}
\vspace{-0.5em}
\begin{tabular}{l c r r r} 
\toprule
& & \hspace{-8mm}Columbia & Carvalho & Korus\\
\midrule
Bondi et al.~\cite{bondi2017cvprw} &  & 0.70 & 0.76 & 0.53 \\
Huh "Camera"~\cite{huh2018forensics} & & 0.70 & 0.73 & 0.15 \\
Huh "Image"~\cite{huh2018forensics} & & 0.97 & 0.75 & 0.58\\
Huh "EXIF"~\cite{huh2018forensics} & & \textbf{0.98} & 0.87 & 0.55 \\
\midrule
 & Patch Size\\
Spectral Gap & 128 & 0.95 & 0.95 & \textbf{0.65} \\
 & 256 & 0.94 & \textbf{0.97} & 0.59\\
Modularity Opt. & 128 & 0.95 & 0.90  & 0.60 \\
 & 256 & 0.92 & 0.95 & 0.57\\
Min. Sim. & 128 & 0.94 & 0.92  & 0.65 \\
 & 256 & 0.89 & 0.96 & 0.59\\
Mean Sim. & 128 & 0.95 & 0.91 & 0.60 \\
 & 256 & 0.92 & 0.95 & 0.57\\
\bottomrule
\end{tabular}
\label{tab:result_benchmark_map}
\end{table}

\begin{table}[t]\centering 
\caption{$P_D$ at $P_{FA}=0.01$ on Benchmark Databases}
\vspace{-0.5em}
\begin{tabular}{l c r r r} 
\toprule
& & \hspace{-8mm}Columbia & Carvalho & Korus\\
\midrule
Bondi et al.~\cite{bondi2017cvprw} &  & 0.07 & 0.03 & 0.01 \\
Huh "EXIF"~\cite{huh2018forensics} &  & 0.51 & 0.17  & 0.01\\
\midrule
 & Patch Size\\
Spectral Gap & 128 & 0.56 & 0.43 & \textbf{0.16} \\
 & 256 & 0.60 & \textbf{0.80} & 0.05\\
Modularity Opt. & 128 & 0.46 & 0.41 & 0.02 \\
 & 256 & \textbf{0.64} & 0.38 & 0.03\\
Min. Sim. & 128 & 0.39 & 0.49 & 0.11 \\
 & 256 & 0.36 & 0.75 & 0.07\\
Mean Sim. & 128 & 0.23 & 0.46 & 0.01 \\
 & 256 & 0.45 & 0.51 & 0.01\\
\bottomrule
\end{tabular}
\label{tab:result_benchmark_pd}
\end{table}

\begin{figure*}
\null\hfill
\begin{subfigure}[t]{0.32\linewidth}
\centering
\includegraphics[width=\linewidth]{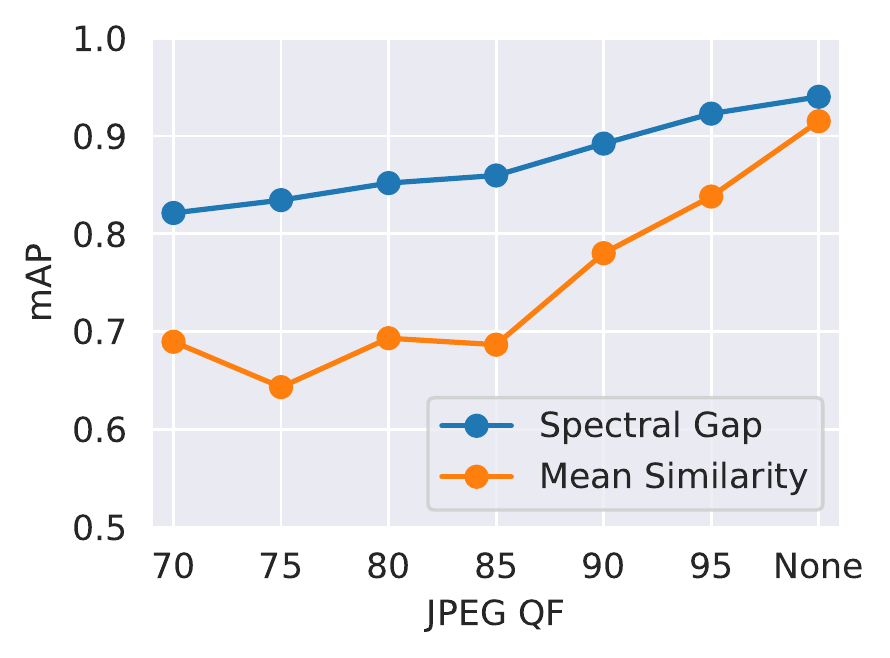}
\caption{Columbia}
\end{subfigure}
\hfill
\begin{subfigure}[t]{0.32\linewidth}
\centering
\includegraphics[width=\linewidth]{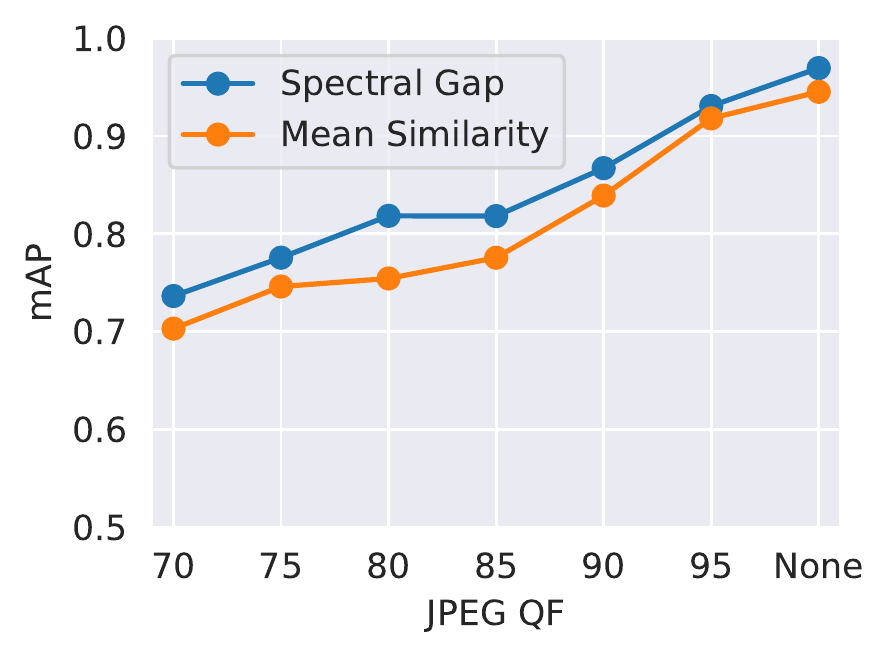}
\caption{Carvalho}
\end{subfigure}
\hfill
\begin{subfigure}[t]{0.32\linewidth}
\centering
\includegraphics[width=\linewidth]{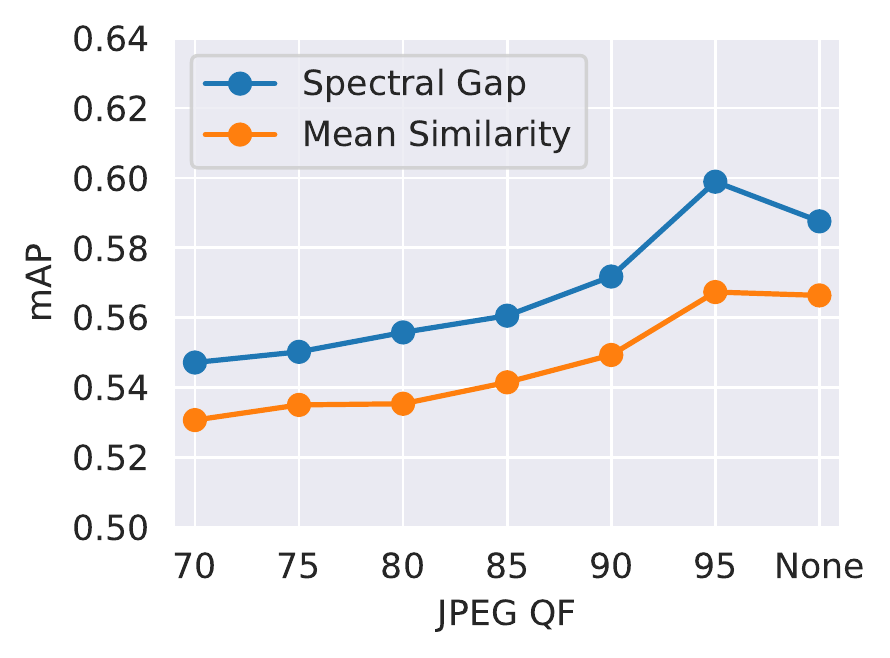}
\caption{Korus}
\end{subfigure}
\hfill\null
\caption{Forgery detection mAP on benchmark images with global JPEG compression applied.}
\label{fig:JPEG_mAP}
\end{figure*}

Table~\ref{tab:result_benchmark_map} shows mAP scores for our proposed approaches as well as the methods we compared against. The mAP measure was chosen to compare to the reported scores in~\cite{huh2018forensics}. All of our proposed approaches achieved higher mAP scores than the exisitng Bondi et al., and Huh ``Camera'' approaches. The Huh ``EXIF'' method achieved a higher mAP score of 0.98 in the Columbia dataset. However, our proposed Spectral Gap method outperformed it on the more challenging Carvalho and Korus datasets, with mAP scores of 0.97 using a 256$\times$256 forensic patch size on Carvalho, and 0.65 using a 128$\times$128 forensic patch size on the Korus dataset.

Table~\ref{tab:result_benchmark_pd} shows the detection rate ($P_D$) at the low false alarm rate of 0.01 for the tested approaches. Operating at low false alarm rates are important for investigators, due to the high cost of falsely identifying authentic content as tampered, and due to the high volume of unaltered content that they encounter. For the Columbia database, the Modularity Optimization approach with a patch size $256\times256$ achieved highest detection rate of 0.64. The Spectral Gap approach achieved highest rates of 0.80 on the Carvalho dataset with a size of $256\times256$, and 0.16 on the Korus dataset with a size of $128\times128$. 

The results of this experiment show that our proposed technique exceeds prior-art forgery detection performance on the three tested benchmark datasets, and significantly outperforms these approaches in challenging datasets. Importantly, the experiment also shows that significant benefit is achieved when using community-structure aware approaches over the community-naive approaches (Mean and Minimum), especially at low false alarm rates. 


Interestingly, we note that while the Modularity Optimization approach performed very well on the Columbia database, performance degraded in the other databases. The forged regions in Columbia dataset are much larger relative to the size of the image than in the Carvalho and Korus datasets. One possible reason behind this is performance decrease is that modularity optimization techniques are known to have difficulties detecting communities that are small relative to the size of the graph~\cite{fortunato2007resolution}, referred to as the ``resolution problem.'' 
\begin{figure*}
\null\hfill
\begin{subfigure}[t]{0.31\linewidth}
\centering
\includegraphics[width=\linewidth]{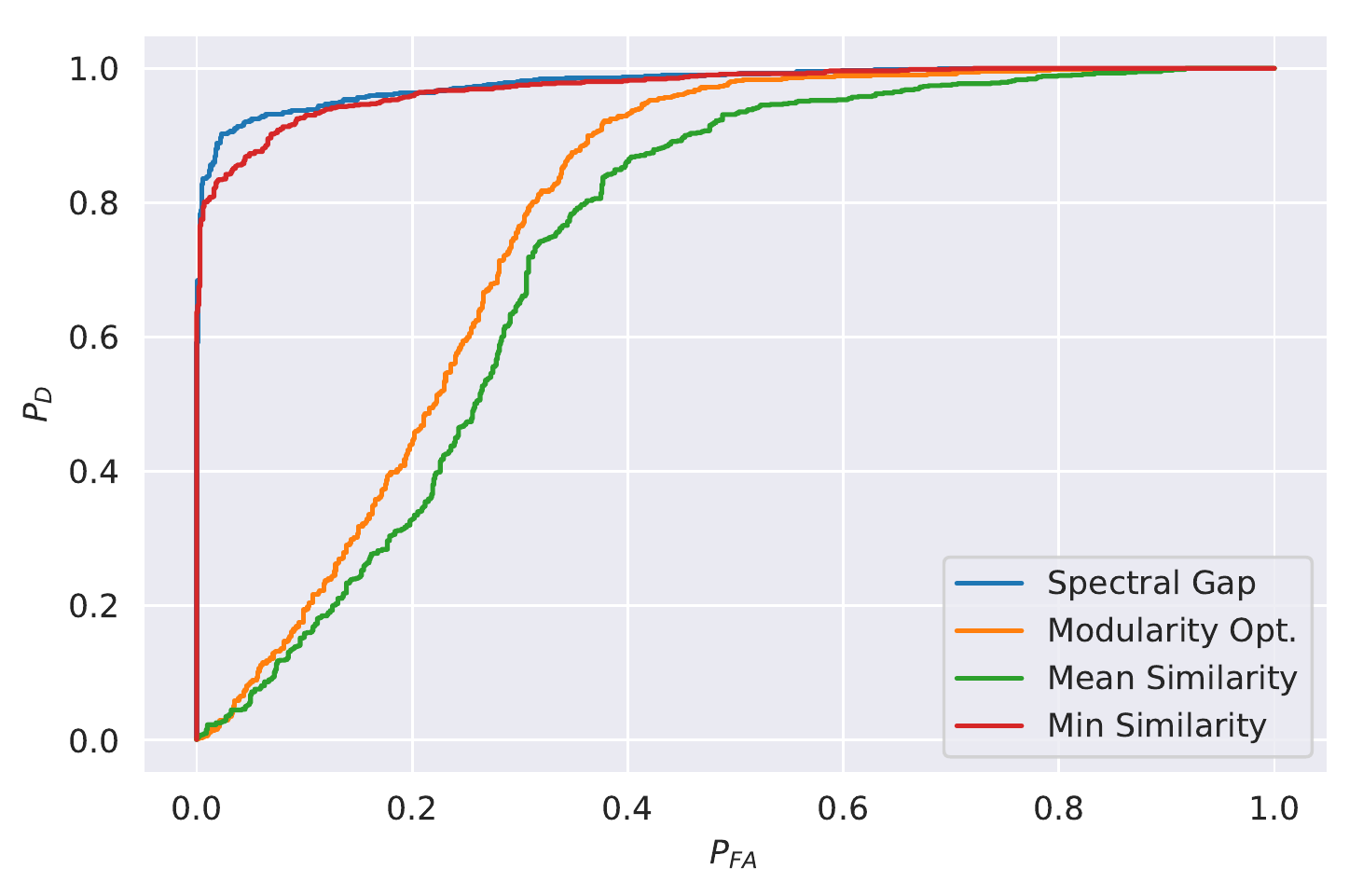}
\caption{ROC, $256 \times 256$ Forgery Size}
\end{subfigure}
\hfill\
\begin{subfigure}[t]{0.31\linewidth}
\centering
\includegraphics[width=\linewidth]{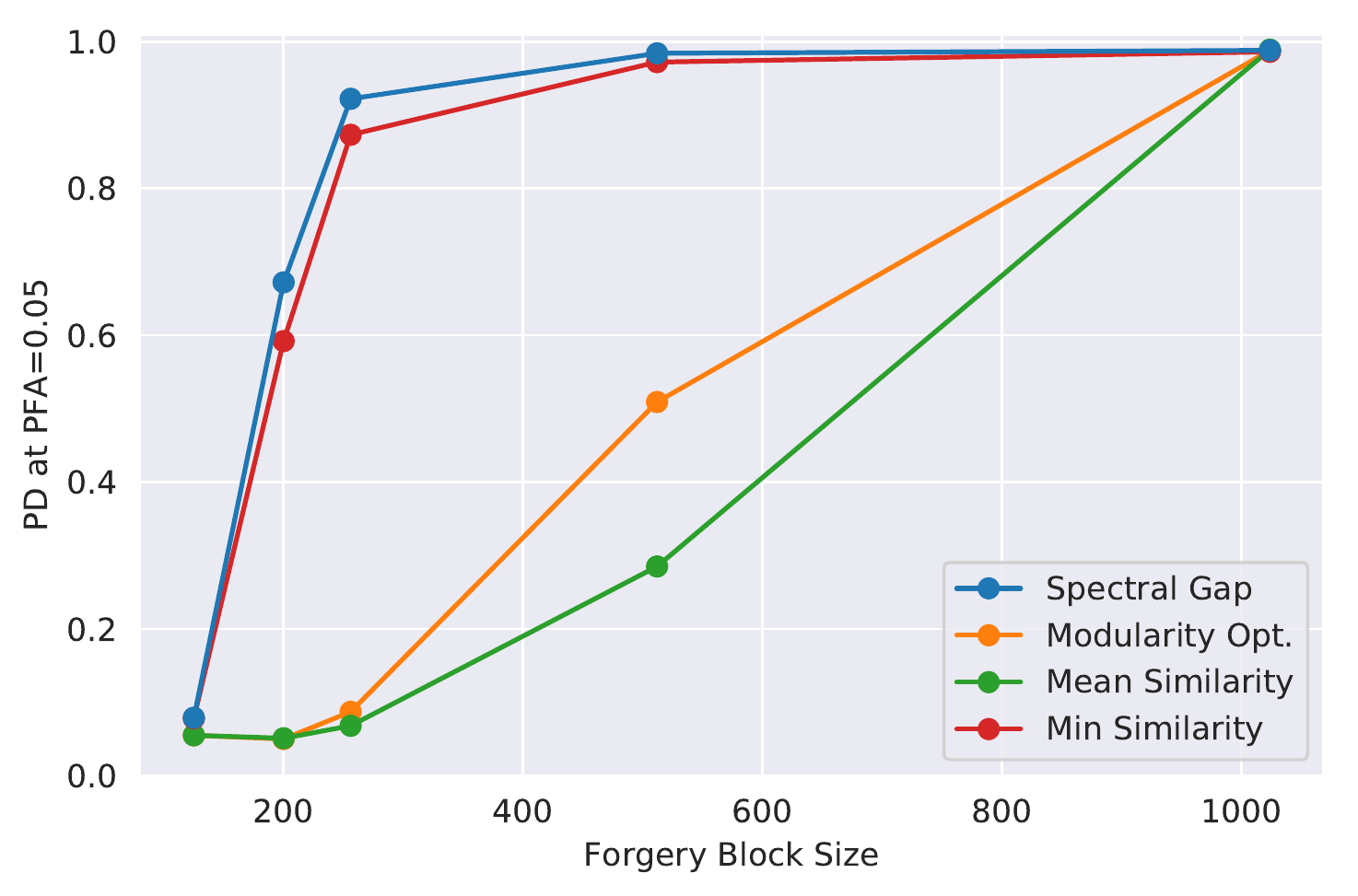}
\caption{$P_D$ at $P_{FA}=0.05$ by Forgery Size}
\end{subfigure}
\hfill
\begin{subfigure}[t]{0.36\linewidth}
\centering
\includegraphics[width=\linewidth]{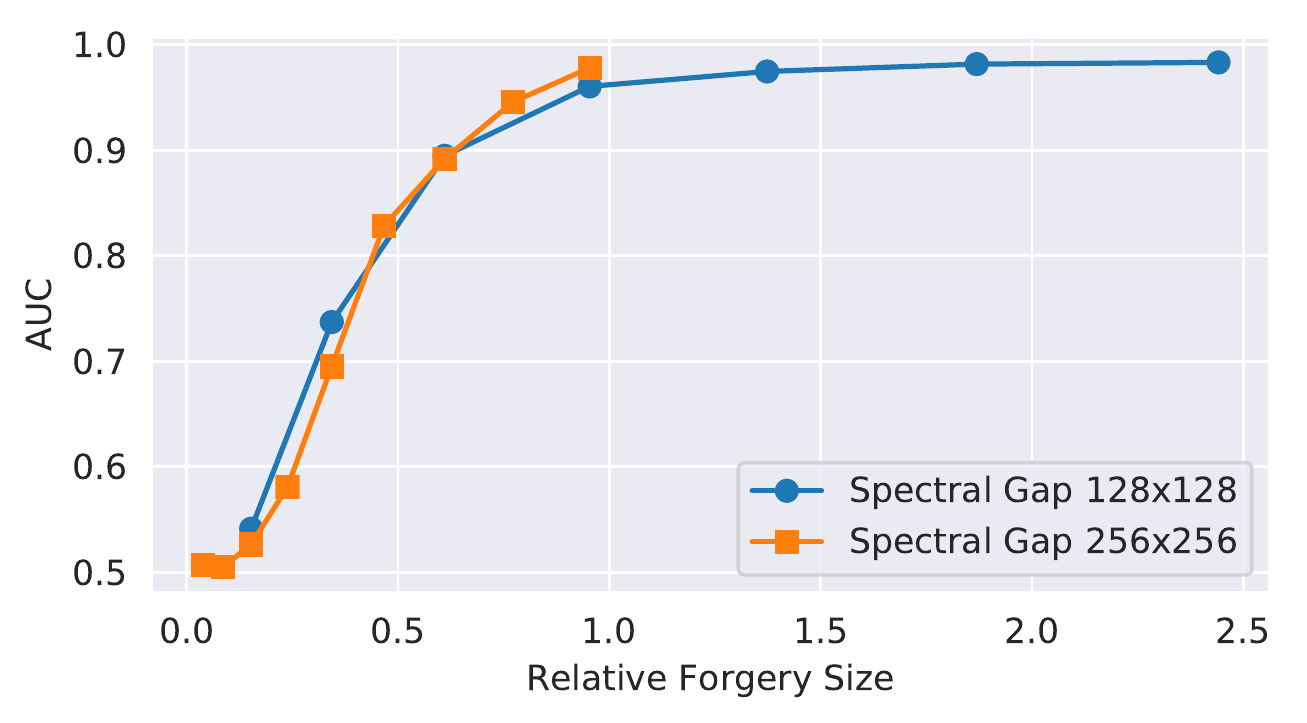}
\caption{AUC by Relative Forgery Size}
\end{subfigure}
\hfill\null
\caption{Synthetic forgery detection results. The plot (a) shows ROC curves for proposed and naive methods in synthetic forgeries with forgery block size of 256$\times$256. 
Plot (b) shows the forgery detection probability for each method for different forgery block sizes. Plot (c) shows the ROC area-under-the-curve at challenging forgery block sizes using the spectral gap method with two different forensic patch sizes. In (c) forgery block size is displayed as relative area to the forensic patch size.\looseness=-1}
\label{fig:vtl_result_1}
\end{figure*}

\textit{Effects of global post processing}:
Here we investigated the impact of additional global post processing on forgery detection performance. To do this, we repeated the above forgery experiment, with additional JPEG compression applied to each image, controlling for JPEG compression quality factor (QF). For each image, with a specified QF, we then computed its respective forensic similarity matrix using a forensic similarity model trained on images with JPEG re-compression applied of that QF. Finally, we computed the community-aware Spectral Gap and community-naive Mean Similarity forgery detection measures. We investigated performance at \mbox{$\text{QF} \in \left\lbrace 70, 75, 80, 85, 90, 95 \right\rbrace$}, where lower quality factors correspond to heavier compression effects.

Forgery detection performance under global JPEG compression is shown in Fig.~\ref{fig:JPEG_mAP}. As expected forgery detection performance decreases with heaver compression (lower QF). Interestingly, the community-aware Spectral Gap approach is less affected by JPEG compression than the community-naive Mean Similarity measure in the Columbia dataset. Another interesting observation is that performance decrease is steeper in the Carvalho dataset than in the Columbia dataset. Two possible explanations are that the salient features in the Carvalho dataset are more affected by JPEG compression, or that the relatively smaller size of the forged regions are harder to detect under significant compression.

\subsection{Characterizations on Synthetic Forgeries}
In the above experiment, we observed that some forgeries were more challenging to detect than others. 
In this experiment, we studied the impacts of the tampered region size and patch size on forgery detection performance. To do this, we started with a database of images from the VISION database~\cite{shullani2017vision}, using the ``natural'' images from the 35 cameras in the database. This database was chosen since the camera models in the database were different from those used to train our proposed system, representing a more practical scenario. We then created synthetic forgeries by copying a small-block from one image into another at random locations. We refer to the size of this block as the \textit{forgery block size}. We tested detection rates at different forgery block sizes, and for a given forgery block size, we created 1000 such forgeries. To control for the size of the image, we ensured that the host image was of size 3264$\times$ 2448, the most common size in the dataset. For unaltered images, we used 1000 images randomly chosen of size 3264$\times$2448. To ensure an appropriate similarity signal was present, the donor and host images were captured by different camera models. We then calculated the proposed detection measures on each forged and authentic image to evaluate performance.

Fig.~\ref{fig:vtl_result_1} shows forgery detection performance for the two community approaches and two naive approaches, using an input patch size of 256$\times$256. 
Fig.~\ref{fig:vtl_result_1}(a) shows ROC curves for these approaches at the $256\times 256$ forgery block size. In this case, the Spectral Gap and Minimum Similarity methods demonstrated improved performance over the Modularity Optimization and Mean Similarity approaches. Notably, the Spectral Gap showed significant improvement over the community-naive Minimum Similarity method at low false alarm rates between 0 and 0.10.

Fig.~\ref{fig:vtl_result_1}(b) shows the positive detection rates of the four approaches at a false alarm rate of 0.05. In general, all approaches performed poorly at the smallest forgery size of $128\times 128$, and improved with forgery block size. All approaches achieved $P_D > 0.98$ with the forgery block size of $1024\times 1024$. This experiment demonstrates that forgery size has a significant effect on detection performance.

In particular, forgery size impacted the Modularity Optimization method. Modularity optimization methods are well known to have difficulty in detecting communities much smaller in scale to the size of the graph~\cite{fortunato2007resolution}. This is also logical for the Mean Similarity method, as the mean value is inherently tied to the percentage of tampered patches relative to the whole image, likely achieving highest detection performance when 50\% of the image is tampered. The largest forgery block size of $1024\times 1024$ is approximately 13\% of the image size, whereas $256\times 256$ is less than 1\% of the image size. This result suggests that the Modularity Optimization and Mean Similarity results are sensitive to the size of the forged region relative to the size of the image, whereas the Spectral Gap method is only sensitive to forgery size up to the forensic patch size, as we discuss in more detail below. Further experimentation on the impact of forgery size relative to the image size may lend further insight on this finding.

We further experimented using the Spectral Gap method at more challenging forgery block sizes. Fig.~\ref{fig:vtl_result_1}(c) shows the ROC area-under-the-curve (AUC) of forgery block sizes from $50\times50$ to $250\times 250$, for forensic patch sizes of $128\times 128$ and $256\times 256$. 
We normalized the x-axis of the result relative to the forgery size, equal to the area of the forgery size divided by the area of the analysis patch size. We see that Spectral Gap performance steeply improves for forgery sizes larger than the analysis patch size. For both forensic patch sizes, forgeries smaller than the size of the forensic similarity patch were less likely to be detected. As a result, we would expect smaller patch sizes to improve performance in small forgery scenarios. This result is consistent with results on the benchmark datasets; methods using a patch size of $256\times256$ performed better on the Columbia dataset, which generally have large forged regions, and methods with patch sizes of $128\times128$ generally performed better on Korus dataset, which has smaller forged regions.

The results of this experiment highlight strong dependence of the proposed approaches on the size of the tampered region. The Spectral Gap method performed well when forgery sizes were equal to or greater than the analysis patch size. At challenging forgery sizes, the Spectral Gap method showed performance improvements over the naive Minimum Similarity approach, demonstrating the importance of identifying community structures captured by the Forensic Similarity Graph. The Modularity Optimization method worked well only when the forgery size was large, demonstrating a dependence on size of the tampered region relative to the size of whole image for this approach.

\begin{figure*}

\begin{minipage}[c]{0.025\linewidth}~~~\end{minipage}
\begin{minipage}[c]{0.975\linewidth}\includegraphics[trim={0 0.1in 0 0.95in},clip,width=\linewidth]{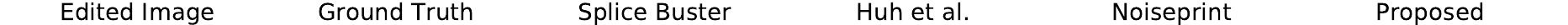}
\end{minipage}

\begin{minipage}[c]{0.025\linewidth}(a)\end{minipage}
\begin{minipage}[c]{0.975\linewidth}\includegraphics[width=\linewidth]{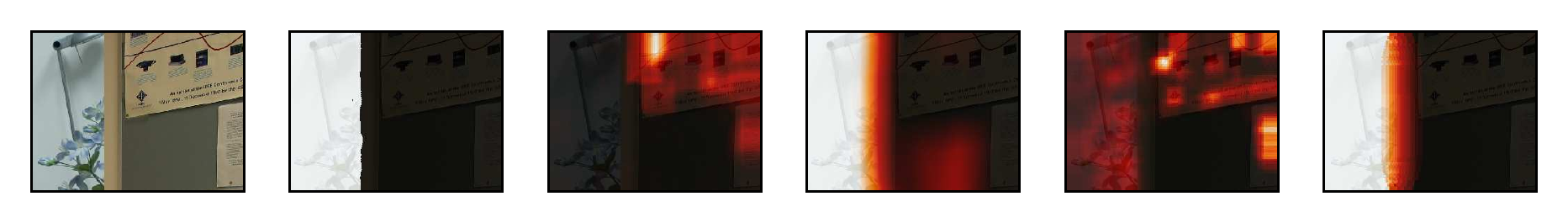}
\end{minipage}

\begin{minipage}[c]{0.025\linewidth}(b)\end{minipage}
\begin{minipage}[c]{0.975\linewidth}\includegraphics[width=\linewidth]{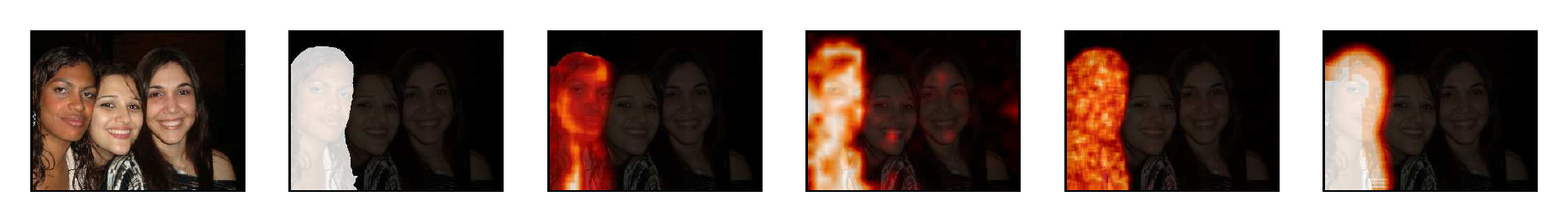}
\end{minipage}

\begin{minipage}[c]{0.025\linewidth}(c)\end{minipage}
\begin{minipage}[c]{0.975\linewidth}\includegraphics[width=\linewidth]{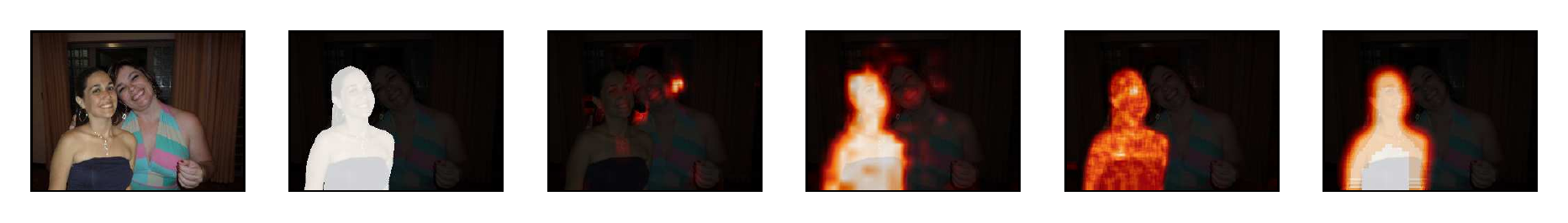}
\end{minipage}

\begin{minipage}[c]{0.025\linewidth}(d)\end{minipage}
\begin{minipage}[c]{0.975\linewidth}\includegraphics[width=\linewidth]{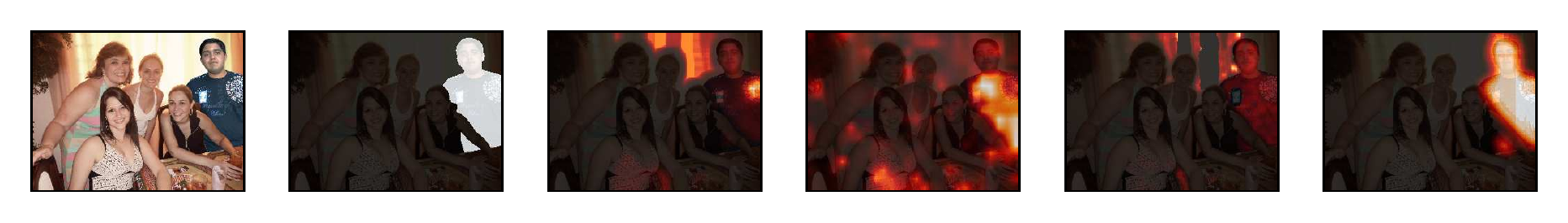}
\end{minipage}

\begin{minipage}[c]{0.025\linewidth}(e)\end{minipage}
\begin{minipage}[c]{0.975\linewidth}\includegraphics[width=\linewidth]{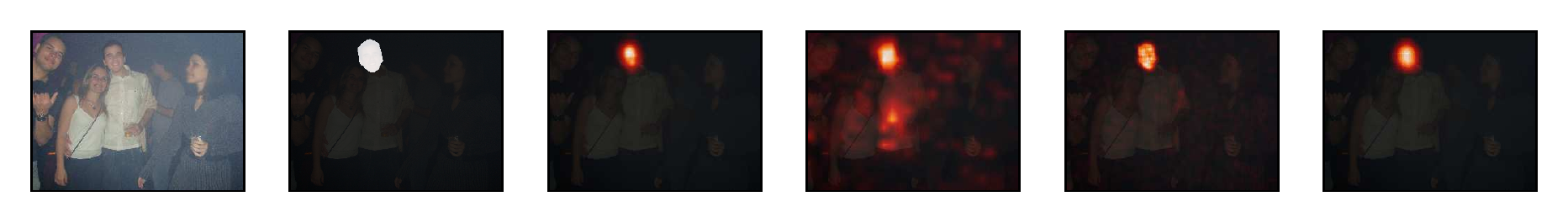}
\end{minipage}

\begin{minipage}[c]{0.025\linewidth}(f)\end{minipage}
\begin{minipage}[c]{0.975\linewidth}\includegraphics[width=\linewidth]{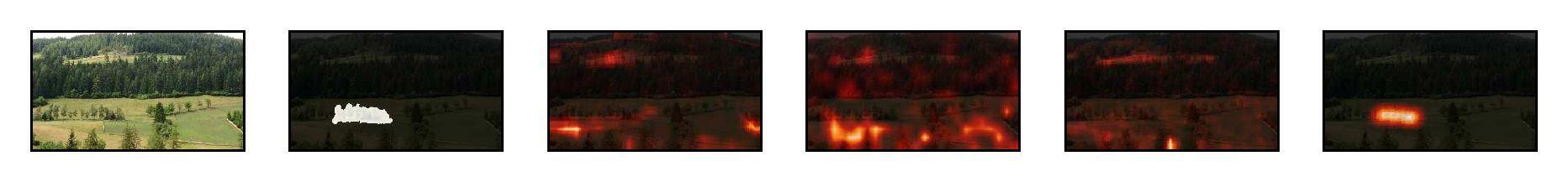}
\end{minipage}

\begin{minipage}[c]{0.025\linewidth}(g)\end{minipage}
\begin{minipage}[c]{0.975\linewidth}\includegraphics[width=\linewidth]{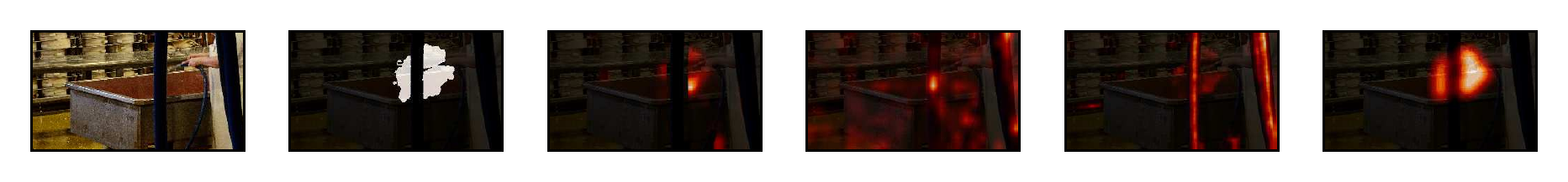}
\end{minipage}

\begin{minipage}[c]{0.025\linewidth}(h)\end{minipage}
\begin{minipage}[c]{0.975\linewidth}\includegraphics[width=\linewidth]{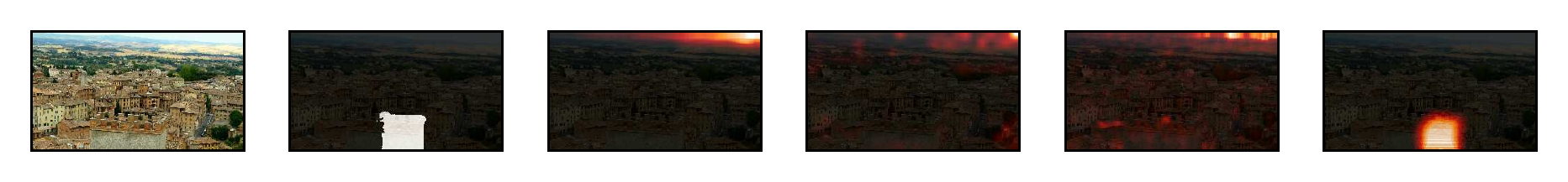}
\end{minipage}

\begin{minipage}[c]{0.025\linewidth}(i)\end{minipage}
\begin{minipage}[c]{0.975\linewidth}\includegraphics[width=\linewidth]{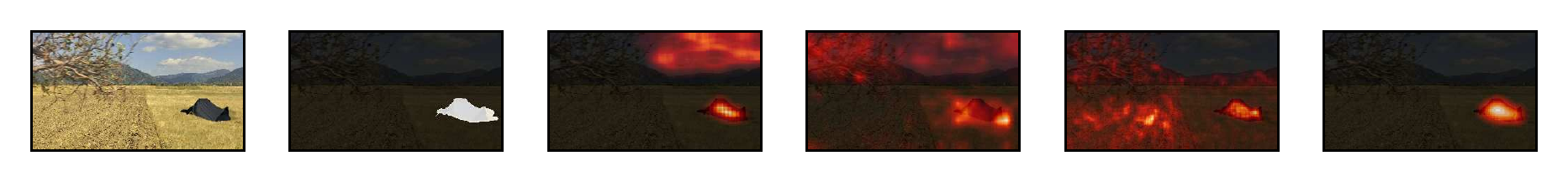}
\end{minipage}

\begin{minipage}[c]{0.025\linewidth}~~~\end{minipage}
\begin{minipage}[c]{0.975\linewidth}\includegraphics[trim={0 0.1in 0 0.95in},clip,width=\linewidth]{figs/loc_comparison2/carvalho_sc_splicing-04_labels.pdf}
\end{minipage}

\caption{Localization examples from the benchmarking datasets. Localization is performed using comparison methods, and our proposed Spectral Clustering method. Example images are from the (a) Columbia, (b-e) Carvalho, and (f-i) Korus datasets.}
\label{fig:loc_db_examples}
\end{figure*}

\section{Experimental Results: Forgery Localization}
\label{sec:localization_experiments}

In this set of experiments, we tested the efficacy of our proposed forgery localization approaches. To do this, we performed forgery localization on tampered images from the three publicly available datasets used in the prior experiments. We evaluated forgery localizations using a variety of scoring measures, and compared to existing-art methods. The methods we compare to include localization methods by Bondi et al.~\cite{bondi2017cvprw} and Huh et al.~\cite{huh2018forensics}, which were compared against in the previous forgery detection experiments. In addition, we compared to two localization-only techniques, the rich-model based SpliceBuster~\cite{cozzolino2015splicebuster} and deep-learning based NoisePrint~\cite{cozzolino2018noiseprint} methods by Cozzolino et al. The results of our experiments shows that our proposed approaches outperform these existing-art methods in the majority of scenarios.

To construct the forensic similarity graph for each forged image, we first sampled the image using patches of size $128\times128$, with 75\% overlap in each dimension. The smaller patch size and higher overlap was chosen to increase spatial resolution. Then, we calculated the forensic similarity between image patches using the deep-learning approach described in~\cite{mayer2019similarity}, trained on four million image patches from 80 camera models. The forensic similarity graph was constructed for each forged image as described in Sec.~\ref{sec:approach}. Then, forgery localization was performed according to the proposed methods in Sec.~\ref{sec:approach:ssec:detection}. For the Spectral Clustering methods, we tested both cases with the non-normalized Laplacian, and the normalized Laplacian. For the Modularity Optimization approach, we used a edge threshold $t=0.7$. Finally, the pixel-level forgery prediction map was created according to the method in Sec.~\ref{sec:approach:ssec:pixel_level} using a Gaussian smoothing function with window size of $32\times32$.

\begin{figure}

\begin{minipage}[c]{0.035\linewidth}~~~\end{minipage}
\begin{minipage}[c]{0.955\linewidth}\includegraphics[width=\linewidth]{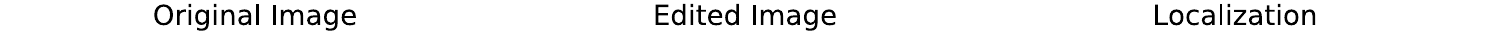}
\end{minipage}

\begin{minipage}[c]{0.035\linewidth}(a)\end{minipage}
\begin{minipage}[c]{0.955\linewidth}\includegraphics[width=\linewidth]{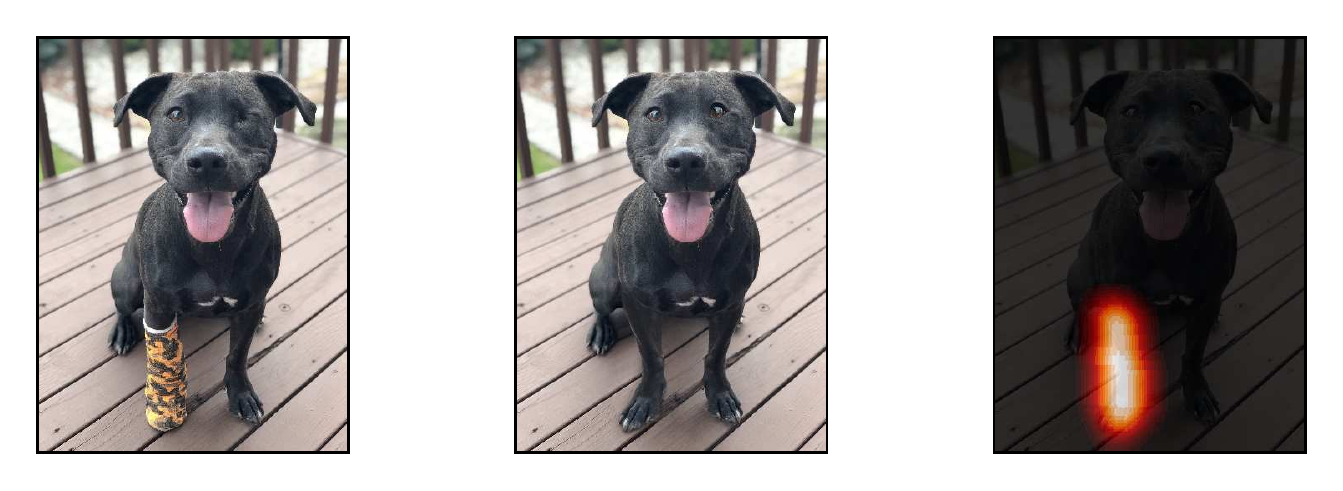}
\end{minipage}

\begin{minipage}[c]{0.035\linewidth}(b)\end{minipage}
\begin{minipage}[c]{0.955\linewidth}\includegraphics[width=\linewidth]{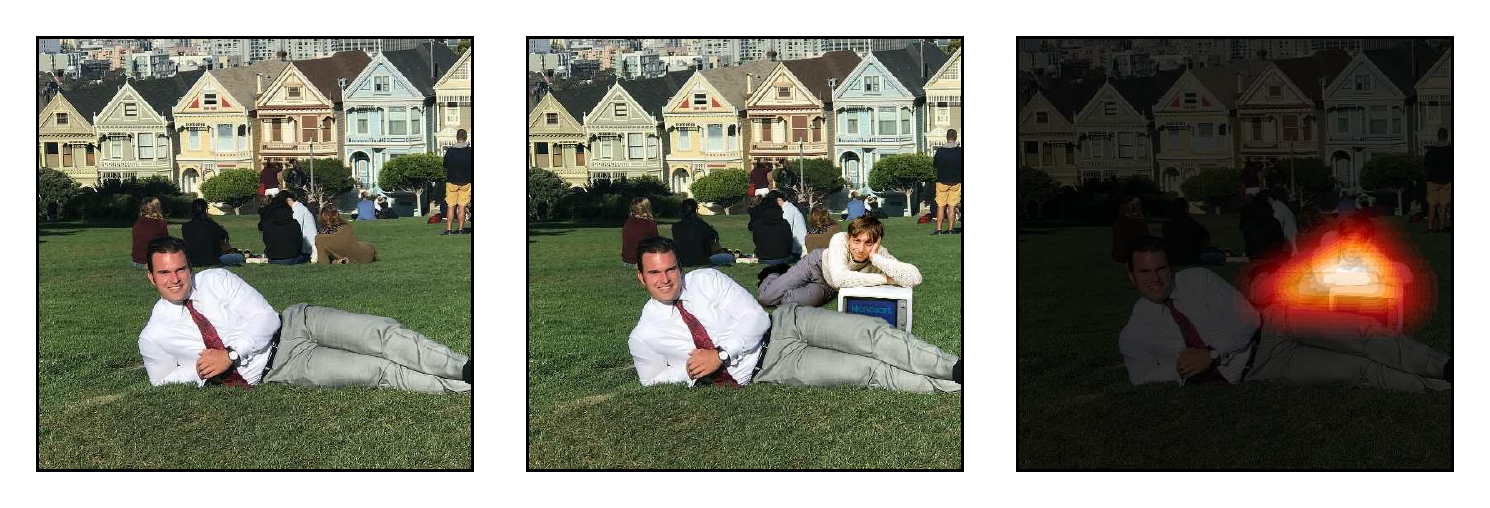}
\end{minipage}


\begin{minipage}[c]{0.045\linewidth}(c)\end{minipage}
\begin{minipage}[c]{0.955\linewidth}\includegraphics[width=\linewidth]{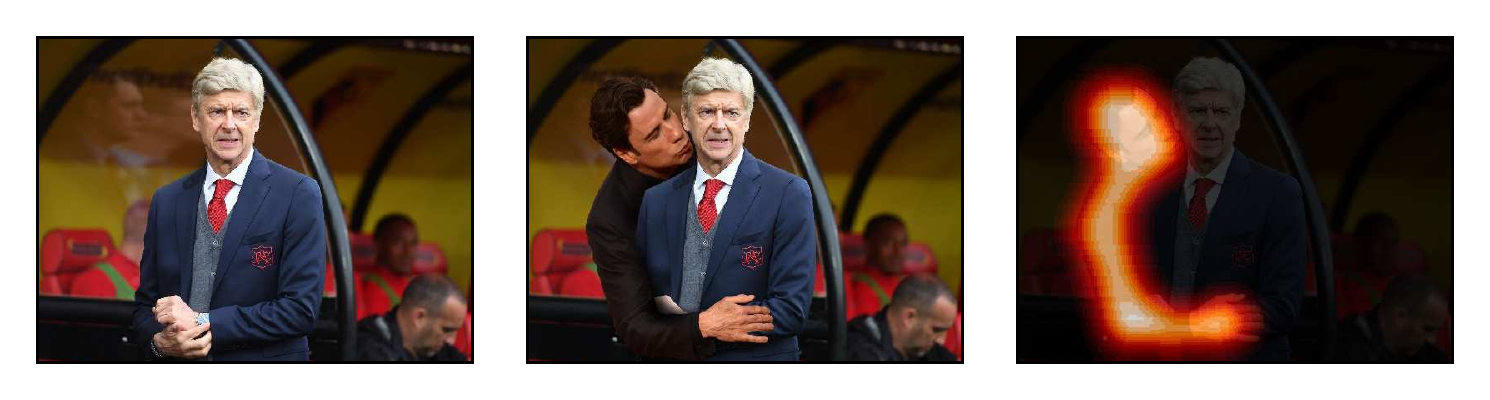}
\end{minipage}

\begin{minipage}[c]{0.045\linewidth}(d)\end{minipage}
\begin{minipage}[c]{0.955\linewidth}\includegraphics[width=\linewidth]{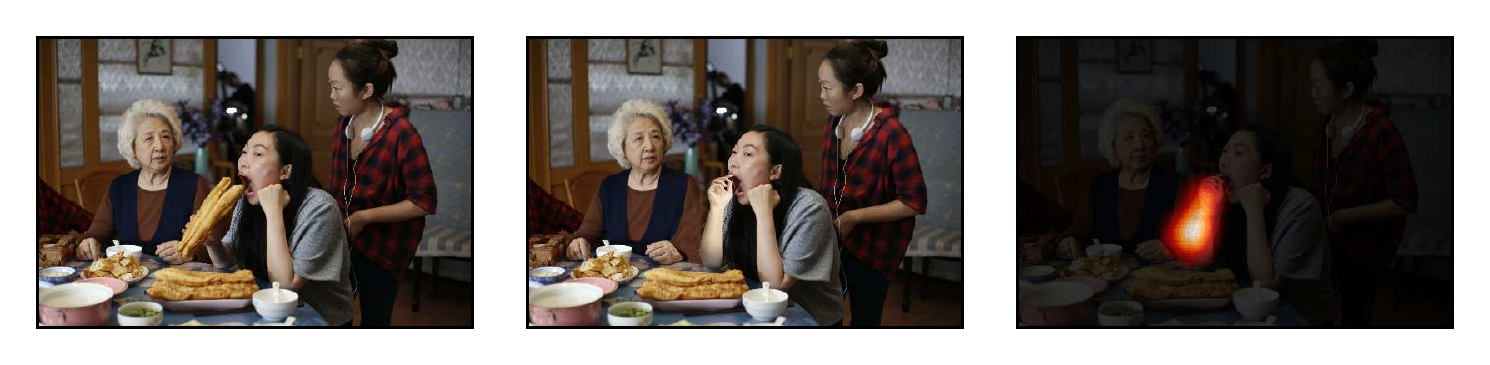}
\end{minipage}


\begin{minipage}[c]{0.045\linewidth}~~~\end{minipage}
\begin{minipage}[c]{0.955\linewidth}\includegraphics[width=\linewidth]{figs/loc_comparison/reddit_labels.pdf}
\end{minipage}
\caption{Spectral clustering localization examples on images downloaded from the social media website Reddit.com. Image editing credit to Reddit users ``artunitinc'' (a), ``ene\_due\_rabe'' (b-c), and ``Hordon\_Gayward'' (d).}
\label{fig:loc_reddit_examples}
\vspace{-1.0em}
\end{figure}

\newcommand{\fone}{$\text{F}_1$ }

\subsection{Localization Performance}

Fig.~\ref{fig:loc_db_examples} shows un-thresholded localization examples on tampered images from the benchmark databases. From left to right, we show the edited image, the ground truth mask, and the localization results from the SpliceBuster algorithm~\cite{cozzolino2015splicebuster}, the Huh et al. algorithm~\cite{huh2018forensics}, the Noiseprint algorithm~\cite{cozzolino2018noiseprint}, and our proposed Spectral Clustering algorithm. Row (a) corresponds to an example tampered image from the Columbia dataset, rows (b)-(e) correspond to tampered images from the Carvalho dataset, and rows (f)-(i) correspond to tampered images from the Korus dataset. These examples show the power of our proposed approach. In many cases, our proposed approach accurately localized the tampered region, even in scenarios where the other algorithms had difficulty such as in the Korus dataset.

In addition, in Fig.~\ref{fig:loc_reddit_examples} we show localization results from images downloaded from a photo-manipulating competition forum on Reddit.com, a social media website. These images contain highly varying and complex tampering techniques. Still, our proposed Spectral Clustering localization algorithm accurately localized the tampered regions of the images. These examples show that our proposed technique is effective on realistically tampered images, including those downloaded from social media websites.

To quantify localization performance, we used several scoring measures. Namely, we used Matthews Correlation Coefficient\footnote{$MCC = \frac{TP \times TN - FP \times FN}{\sqrt{\left(TP+TP\right)\left(TP-FN\right)\left(TN+FP\right)\left(TN+FN\right)}}$} (MCC), the \fone score\footnote{$F_1 = 2\frac{\text{precision}\times\text{recall}}{\text{precision}+\text{recall}}$}, and the Area Under the Curve (AUC) of the reciever operatoring characteristic curve (ROC). The MCC and \fone scores require a choice of threshold, we evaluated both cases where the threshold was chosen for each image, according to the approach in~\cite{huh2018forensics}, and where the threshold was chosen for the entire database. Setting a single threshold for the entire database is a more realistic scenario, since the forensic investigator does not have ground truth data available for each image under investigation and must choose a threshold using outside information. The MCC and \fone scores were calculated for each image and then averaged for each benchmark database, the AUC score was calculated using the ROC determined by all pixels in the database.

\begin{table}
\center
\caption{MCC, per image threshold}
\begin{tabular}{l c c c}
\toprule
 & Columbia & Carvalho & Korus \\
 \midrule
Splice Buster~\cite{cozzolino2015splicebuster} & 0.68 & 0.56 & 0.34 \\
Bondi et al.~\cite{bondi2017cvprw} & 0.50 & 0.35 & 0.22 \\
Huh et al. EXIF~\cite{huh2018forensics} & \textbf{0.87} & 0.46 & 0.23 \\
NoisePrint~\cite{cozzolino2018noiseprint} & 0.74 & 0.72 & 0.31 \\
\midrule
Modularity Opt. (t=0.7) & 0.85 & 0.72 & 0.26 \\
Spectral  Clustering & 0.86 & \textbf{0.80} & \textbf{0.38}\\
Normed Spectral Clust. & 0.84 & 0.77 & 0.25 \\
\bottomrule
\end{tabular}
\label{tab:mcc_im}
\end{table}

\begin{table}
\center
\caption{MCC, per database threshold}
\begin{tabular}{l c c c}
\toprule
 & Columbia & Carvalho & Korus \\
 \midrule
Splice Buster~\cite{cozzolino2015splicebuster} & 0.63 & 0.53 & 0.28 \\
Bondi et al.~\cite{bondi2017cvprw} & 0.45 & 0.29 & 0.14 \\
Huh et al. ``EXIF"~\cite{huh2018forensics}& 0.78 & 0.38 & 0.17 \\
NoisePrint~\cite{cozzolino2018noiseprint} & 0.68 & 0.67 & 0.24 \\
\midrule
Modularity Opt. (t=0.7) & 0.80 & 0.65 & 0.19 \\
Spectral  Clustering & \textbf{0.82} & \textbf{0.75} & \textbf{0.31} \\
Normed Spectral Clust. & 0.79 & 0.70 & 0.20 \\
\bottomrule
\end{tabular}
\label{tab:mcc_db}
\end{table}

\begin{table}
\center
\caption{F1, per image threshold}
\begin{tabular}{l c c c}
\toprule
 & Columbia & Carvalho & Korus \\
 \midrule
Splice Buster~\cite{cozzolino2015splicebuster} & 0.78 & 0.62 & 0.36 \\
Bondi et al.~\cite{bondi2017cvprw} & 0.64 & 0.44 & 0.24 \\
Huh et al. ``EXIF"~\cite{huh2018forensics}& \textbf{0.90} & 0.53 & 0.25 \\
NoisePrint~\cite{cozzolino2018noiseprint} & 0.81 & 0.75 & 0.32 \\
\midrule
Modularity Opt. (t=0.7) & 0.88 & 0.73 & 0.26 \\
Spectral  Clustering. & 0.89 & \textbf{0.82} & \textbf{0.39} \\
Normed Spectral Clust. & 0.88 & 0.78 & 0.26 \\
\bottomrule
\end{tabular}
\label{tab:f1_im}
\end{table}

\begin{table}
\center
\caption{F1, per database threshold}
\begin{tabular}{l c c c}
\toprule
 & Columbia & Carvalho & Korus \\
 \midrule
Splice Buster~\cite{cozzolino2015splicebuster} & 0.74 & 0.59 & 0.30 \\
Bondi et al.~\cite{bondi2017cvprw} & 0.59 & 0.39 & 0.18 \\
Huh et al. ``EXIF"~\cite{huh2018forensics}& 0.82 & 0.45 & 0.20 \\
NoisePrint~\cite{cozzolino2018noiseprint} & 0.76 & 0.70 & 0.26 \\
\midrule
Modularity Opt. (t=0.7) & 0.83 & 0.66 & 0.20 \\
Spectral Clustering & \textbf{0.86} & \textbf{0.77} & \textbf{0.32} \\
Normed Spectral Clust. & 0.83 & 0.70 & 0.20 \\
\bottomrule
\end{tabular}
\label{tab:f1_db}
\end{table}

\begin{table}
\center
\caption{Area Under the Curve (AUC)}
\vspace{0em}
\begin{tabular}{l c c c}
\toprule
 & Columbia & Carvalho & Korus \\
 \midrule
Splice Buster~\cite{cozzolino2015splicebuster} & 0.79 & 0.68 & 0.56 \\
Bondi et al.~\cite{bondi2017cvprw} & 0.77 & 0.64 & 0.60 \\
Huh et al. ``EXIF"~\cite{huh2018forensics}& \textbf{0.96} & 0.72 & 0.57 \\
NoisePrint~\cite{cozzolino2018noiseprint} & 0.85 & 0.76 & 0.58 \\
\midrule
Modularity Opt. (t=0.7) & 0.95 & 0.94 & \textbf{0.73} \\
Spectral  Clustering & 0.93 & 0.89 & 0.60 \\
Normed Spectral Clust. & 0.95 & \textbf{0.97} & 0.72 \\
\bottomrule
\end{tabular}
\label{tab:loc_auc}
\end{table}

Table~\ref{tab:mcc_im} shows MCC scores for our proposed and comparison approaches on the three benchmark datasets, using per-image thresholds, meaning that a different threshold was chosen for each image. For the Columbia dataset, our proposed methods achieved high scores of 0.84 and above, with the Spectral Clustering method performing the highest of 0.86 among the proposed methods, and just under the score achieved by Huh et al. For the more challenging Carvalho dataset, all three of our proposed methods outperformed the comparison methods. The Spectral Clustering algorithm achieved an MCC score of 0.80, which is significantly higher than the next highest comparison method. For the most challenging Korus dataset, our proposed spectral clustering achieved the highest MCC score of 0.38.

Table~\ref{tab:mcc_db} shows MCC scores using per-database thresholds, meaning that a single threshold was used for each database and is a more practical scenario. In all three benchmark datasets, our proposed Spectral Clustering algorithm achieved the highest scores. In the Columbia dataset, all three of our proposed methods outperform the Huh et al. method, suggesting that our proposed technique is more consistent and does not require threshold tuning for each image.

Similar trends were found for the \fone scores, which are shown with per-image thresholds in Table~\ref{tab:f1_im} and with per-database thresholds in Table~\ref{tab:f1_db}. 
Table~\ref{tab:loc_auc} shows the area under the curve (AUC) scores for each method. In general, the three proposed techniques outperform prior art, especially in the more challenging datasets. We note exceptional performance by the Normed Spectral Clustering approach in the Carvalho dataset with a score of 0.97 (with next highest prior-art score of 0.76), and the Modularity Optimization in the Korus dataset with a score of 0.73 (with next highest prior-art score of 0.60).

The results of this experiment shows that our proposed community detection techniques, and in particular the Spectral Clustering method, consistently outperforms existing art demonstrating the power of our proposed Forensic Similarity Graph based approach. This was shown on several benchmarking forgery databases with a variety of scoring measures. These and the forgery detection experiments highlight the importance of considering community structures when analyzing image forgeries.

\subsection{Localization with Multiple Tampering Communities}

For forgery localization so far we have only been concerned with partitioning the tampered content from the unaltered content. For this reason, we have set the number of localized communities $k=2$, where one community is the unaltered content and the other is the tampered content. This approach has been sufficient since the benchmark datasets typically only contain one tampering. However, forged images may contain multiple tampered regions that have undergone different types of manipulations. In which case, a forensic investigator may also wish to partition the different tampered regions, and can do so by considering $k>2$, as mentioned in Sec.~\ref{sec:approach:ssec:detection}. In this experiment, we considered the case of $k>2$ for Spectral Clustering based localization. To do this, we computed the first $k$ eigenvectors of the graph laplacian matrix, and performed k-means clustering using these eigenevectors as feature vectors to determine the community partitions.

\begin{figure}
\null\hfill
\begin{tabular}{c c}
\includegraphics[width=1.0in]{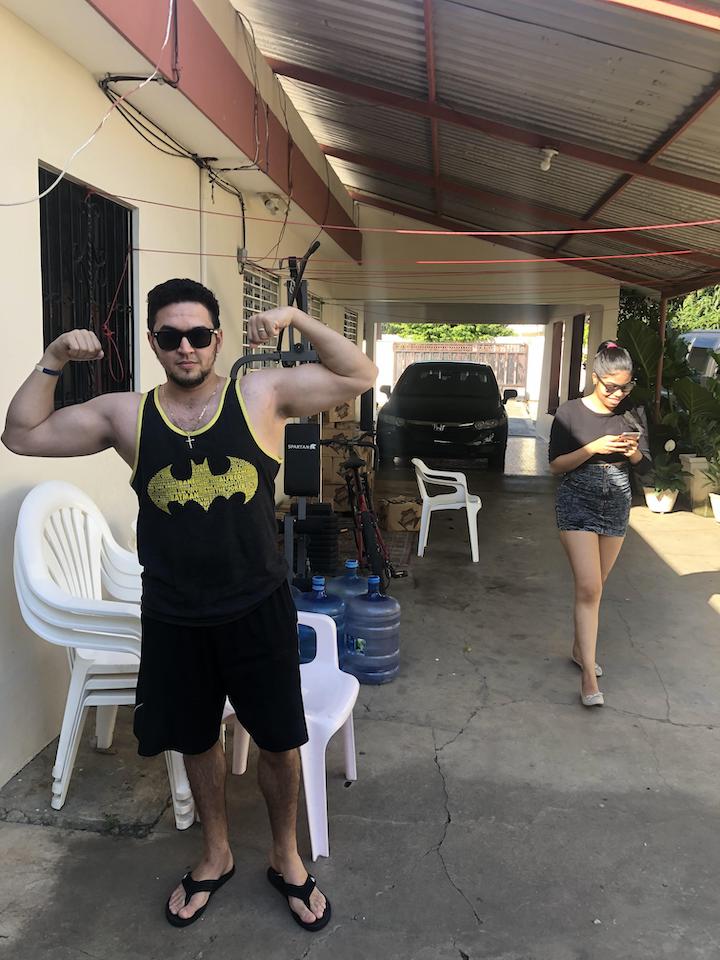} &
\includegraphics[width=1.0in]{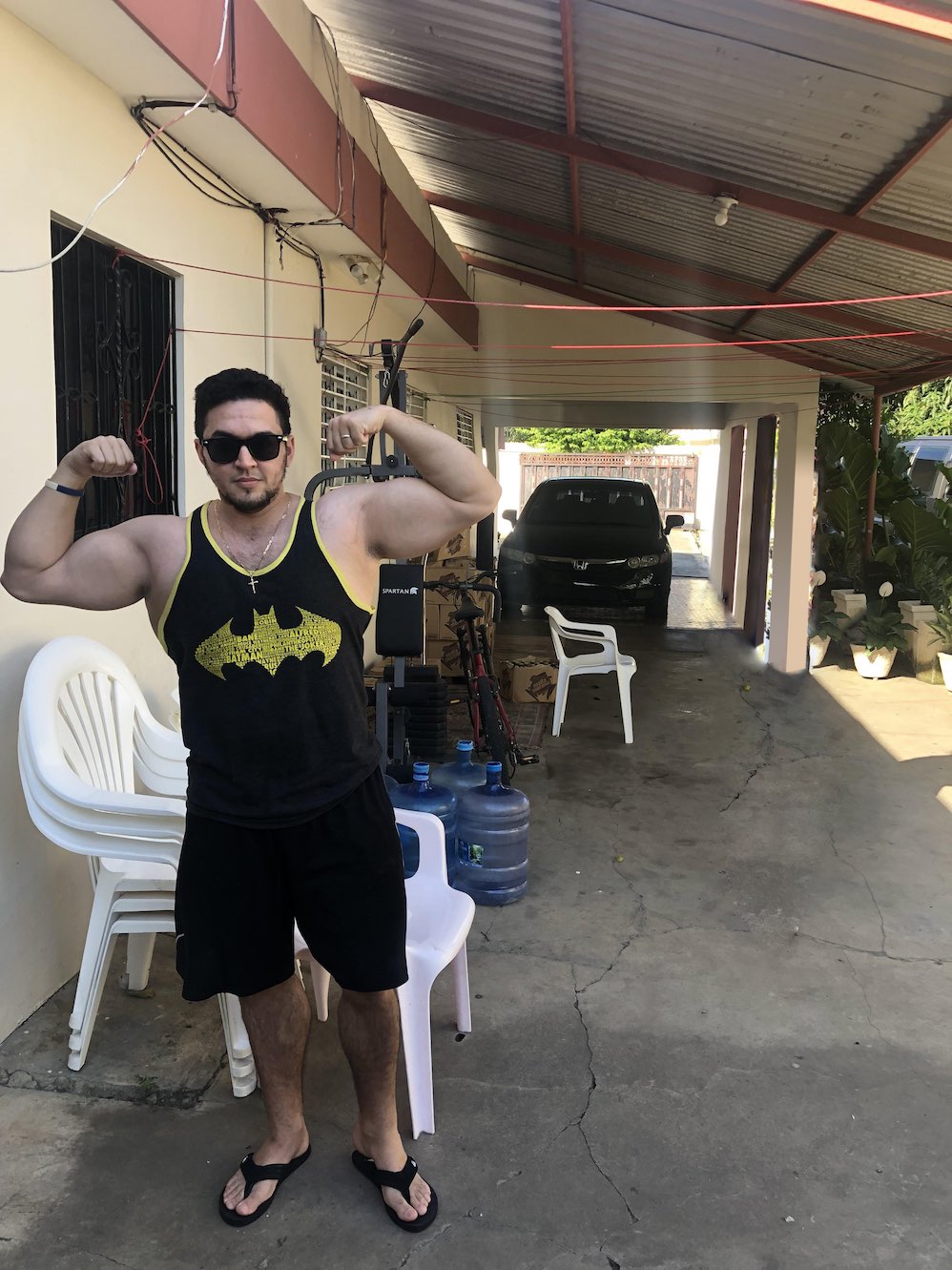}\\
(a) Original & (b) Tampered
\end{tabular}
\hfill\null

\null\hfill
\begin{tabular}{c c c}
\includegraphics[width=1.0in]{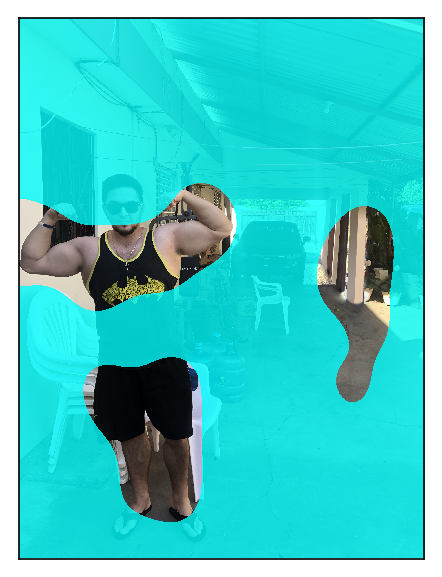} & 
\includegraphics[width=1.0in]{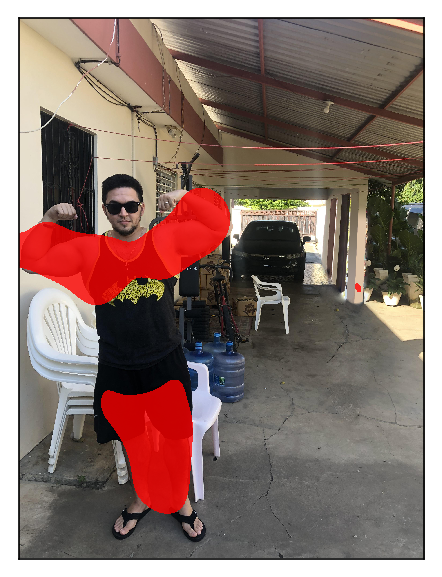} & 
\includegraphics[width=1.0in]{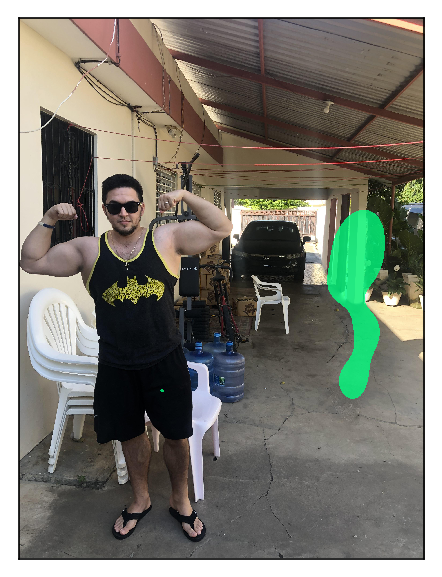} \\
(c) $k=3,c=1$ & (d) $k=3,c=2$ & (e) $k=3,c=3$
\end{tabular}
\hfill\null
\null\hfill
\begin{tabular}{c c}
\includegraphics[width=1.0in]{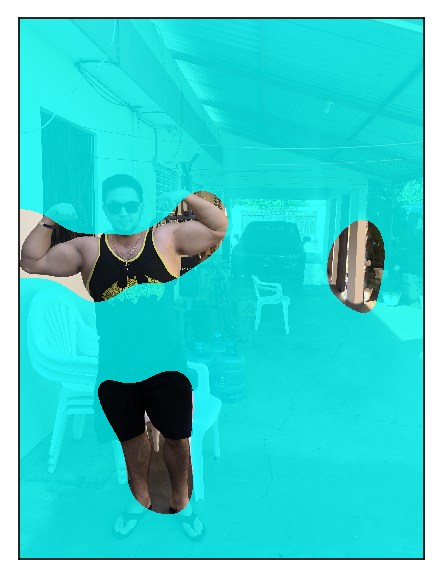} & 
\includegraphics[width=1.0in]{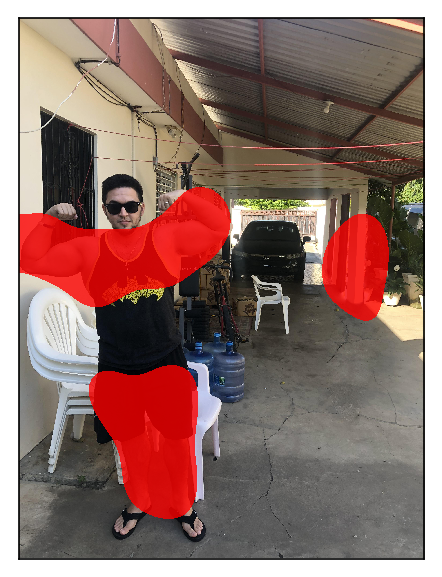} \\
(f) $k=2,c=1$ & (g) $k=2,c=2$
\end{tabular}
\hfill\null
\caption{Forgery localization with multiple tampered regions, each with different processing operations. Comparison shows the effect of the number of localized communities $k$ of 3 and 2. Image editing credit to Reddit user ``Daexsin."}
\label{fig:loc_k3}
\end{figure}

One example of an image with multiple tamperings is shown in Fig.~\ref{fig:loc_k3}. In this example, the forged image shows that 1) the person in the background was removed, and 2) the person in the foreground has their arm and leg muscles enlarged via warping. By setting $k=3$, we were able to differentiate between the communities associated with the tamperings of the arms and legs ($c=2$), and the community associated with the tampering of removing the person in the background ($c=3$). By setting $k=2$, we still differentiated between the tampered regions and the unaltered parts of the image, but not between the two different tamperings. It is important to note that this partitioned the different tampered regions because they were the result of different manipulation processes, and not because they were in different spatial locations. 

Interestingly, we note that by setting $k=2$, we still effectively localized the tampered regions. However, some of the tampered areas of removed-person’s legs were assigned to the “unaltered” community $c=0$. This is likely because those areas were cloned from an unaltered region with minimal additional processing, and as a result were more similar to the unaltered community than the other patches in the tampered community. This result suggests that using $k=2$ is useful for localizing in general, but results may be improved by correctly setting the value for $k$.

The result of this experiment shows that by setting $k>2$, when appropriate, can localize between different types of tampering. Still, setting $k=2$ in the case of multiple tamperings separates the tampered regions from the unaltered regions, and setting the value of $k$ that matches the number of unique tamperings may improve forgery localization.



\section{Computational Complexity}

The execution time of forensic algorithms is an important consideration for the forensic investigator who conducts large scale analyses, such as filtering content at a social media company. To address this, we conducted an experiment to evaluate the computational complexity of the proposed approach. Since the overall approach is composed of several components, we investigated the computation time of each individually. The components are as follows: 1) the CNN-based feature vector extraction step, which is a sub-process of the forensic similarity computation that is particularly computationally demanding~\cite{mayer2019similarity}, performed once per sampled image patch for the $n$ sampled patches, 2) construction of the Forensic Similarity Graph, by computing the forensic similarity matrix, requiring $\frac{n^2 - n}{2}$ forensic similarity calculations, and finally 3) the community detection step, which for the Spectral Clustering approach primarily consists of computing the eigendecomposition of the forensic similarity matrix, which runs in $O(n^3)$ time. 

To conduct this experiment, we started with a standard 1080$\times$1080 pixel image. Then, we computed the elapsed processing time of each algorithm component individually. Since the computational complexity of each step is tied to the number of sampled patches, we investigated different patch sizes with different sampling overlap. Computation of features and forensic similarity for graph construction were performed using Tensorflow v1.15.0, community detection was performed using the spectral clustering approach, with the eigendecomposition computed using NumPy v1.17.4. Experiments were conducted on a single processor computer with Intel i7-8700K CPU clocked at 3.70GHz, and Nvidia GeForce GTX 1080 Ti GPU.\looseness=-1

\begin{table}
\caption{Computation Time}
\setlength{\tabcolsep}{0.5em}
\begin{tabular}{c c c c c c c}
\toprule
& & & \multicolumn{4}{c}{Elapsed Time (s)}\\
\cmidrule{4-7}
Patch Size & Overlap & $n$ & Features & Graph Constr. & Com. Det. & \textbf{Total} \\
\midrule
256  & 50\% & 49 & 0.072 & 0.041 & 0.001 & \textbf{0.115} \\
256  & 75\% & 196 & 0.221 & 0.164 & 0.008 & \textbf{0.393} \\
128  & 50\% & 256 & 0.082 & 0.251 & 0.015 & \textbf{0.348} \\
128  & 75\% & 961 & 0.193 & 1.831 & 0.201 & \textbf{2.225} \\
\bottomrule
\end{tabular}
\label{tab:computation_time}
\end{table}

Elapsed computation time for each step, and total elapsed computation time, is shown in Table~\ref{tab:computation_time}. For a small number of large patches, the feature extraction step dominates computation time. For a large number of small patches, graph construction dominates computation time. For larger images, smaller patch sizes, or more aggressive sample, it may be possible that the community detection step dominates (with the $n^3$ complexity relationship), but we haven't encountered such a scenario in our investigations.

In addition to patch size and number of sampled patches, the hardware has a significant impact on computation time, as well as the architecture of the CNN-based feature extractor. The feature extractor used for similarity calculations in our approach is relatively light-weight, and increasing complexity of the architecture may increase computation time. Still, the results of this experiment show that forgery detection and localization can be achieved on the order of $10^{-1}$ to $10^0$ seconds per image.

\section{Conclusion}
We proposed an abstract, graph-based representation of an image, which we call the Forensic Similarity Graph. This representation can be used to analyze images for evidence localized tampering with high accuracy.
Tampered and unaltered regions form unique structures that align with a concept called ``communities'' in graph-theory literature. As a result, forgery detection is performed by identifying multiple communities, and forgery localization
is performed by partitioning these communities. We experimentally showed that this approach signficantly outperforms naive implementations that do not consider this community structure, including prior art.\looseness=-1

\section{Acknowledgments}
This material is based upon work supported by the National Science Foundation under Grant No. 1553610. Any opinions, findings, and conclusions or recommendations expressed in this material are those of the authors and do not necessarily reflect the views of the National Science Foundation.

The authors would also like to thank Luca Bondi for his assistance with this work.
\bibliographystyle{IEEEtran}
\bibliography{bib/graph.bib,bib/mmf.bib,bib/mmf2.bib,bib/graph2.bib}

\end{document}